\documentclass[prd,preprint,amsmath,amssymb,floatfix,superscriptaddress,nofootinbib]{revtex4-1}
\usepackage{graphicx, epsfig, bm, amsmath}
\usepackage{amsmath, amssymb}
\usepackage{color}
\usepackage{hyperref}
\usepackage{wasysym}
\usepackage{url}
\usepackage{listings}

\newcommand{\ud}{\,\mathrm{d}}

\definecolor{darkgreen}{cmyk}{0.85,0.2,1.00,0.2}

\def\aap{\rm{A\&A}}                
\def\procspie{\rm{Proc.~SPIE}}   
\def\apjs{\rm{ApJS}}               
\def\mnras{\rm{MNRAS}}             
\def\jcap{\rm{J. Cosmology Astropart. Phys.}}



\begin{document}
\preprint{}
\title{Finding structure in the dark: coupled dark energy, weak lensing, and the mildly nonlinear regime}
	\author{Vinicius Miranda} 
\email{vinim@sas.upenn.edu}
\affiliation{Center for Particle Cosmology, Department of Physics and Astronomy,		University of Pennsylvania, Philadelphia, Pennsylvania 19104, USA}
	\author{Mariana Carrillo Gonz\'alez} 
\email{cmariana@sas.upenn.edu}
\affiliation{Center for Particle Cosmology, Department of Physics and Astronomy,		University of Pennsylvania, Philadelphia, Pennsylvania 19104, USA}
    \author{Elisabeth Krause}
\email{lise@slac.stanford.edu}
\affiliation{Kavli Institute for Particle Cosmology and Astrophysics, Stanford University, Stanford, CA 94305, USA}
	\author{Mark Trodden}
\email{trodden@physics.upenn.edu}
\affiliation{Center for Particle Cosmology, Department of Physics and Astronomy,		University of Pennsylvania, Philadelphia, Pennsylvania 19104, USA}	
\date{\today}
\begin{abstract}
We reexamine interactions between the dark sectors of cosmology, with a focus on robust constraints that can be obtained using only mildly nonlinear scales. While it is well known that couplings between dark matter and dark energy can be constrained to the percent level when including the full range of scales probed by future optical surveys, calibrating matter power spectrum emulators to all possible choices of potentials and couplings requires many computationally expensive n-body simulations. Here we show that lensing and clustering of galaxies in combination with the Cosmic Microwave Background (CMB)  are capable of probing the dark sector coupling to the few percent level for a given class of models, using only linear and quasi-linear Fourier modes. These scales can, in principle, be described by semi-analytical techniques such as the effective field theory of large-scale structure. 
\end{abstract}

\maketitle

\section{Introduction}
In the minimal cosmological model - $\Lambda$CDM - dark matter only interacts gravitationally, and dark energy is described by a cosmological constant. While this scenario is consistent with current observations, the existence of theoretical issues such as fine tuning and the coincidence problem, in addition to multiple, but low statistical significance, anomalies may point to the existence of new physics. Some of these anomalies are: the lack of power at large angles in the CMB angular correlation function \cite{Ade:2013nlj,Ade:2015hxq}; the tension between the Planck CMB estimate of $\sigma_8$ and the lower values inferred from weak lensing \cite{Heymans:2012gg,Hildebrandt:2016iqg}, cluster counts \cite{Ade:2015fva}, and redshift-space distortions \cite{Samushia:2012iq}; and the tension among measurements of the Hubble parameter measured at different redshifts \cite{Riess:2016jrr,Ade:2015xua}. Considering the complexity of models arising from our theories of high energy physics, it is prudent to analyze non-minimal models that might address some or all of these anomalous observations. In general, effective field theory dictates that such a description may include an interaction between dark sectors, and the resulting models have been studied and extensively tested  \cite{Amendola:1999er,Bean:2008ac,Bean:2007ny,Bean:2007nx,Koivisto:2005nr,Amendola:2003eq,Amendola:2003wa,Amendola:2002bs,LaVacca:2009yp,Xia:2009zzb,Pettorino:2013oxa}.

After the final upcoming data release from the Planck mission, new information on the background evolution and the formation of structure in the universe will primarily come from optical surveys, such as the Dark Energy Survey (DES) \cite{2005astro.ph.10346T} and the Large Synoptic Survey Telescope \cite{2012arXiv1211.0310L} (LSST), and ground-based CMB observatories sensitive to both the temperature and polarization spectra of the CMB photons \cite{2014SPIE.9153E..1PB,2016ApJS..227...21T,2016arXiv161002743A}. In both of these cases, gravitational lensing plays a significant role and will provide a vast amount of information about how structures have evolved up to the redshift of background galaxies in the case of the optical surveys, and the CMB itself in the case of CMB lensing. 

Both the DES and LSST surveys will provide high-quality data on scales at which gravitational collapse is highly nonlinear. This makes constraining cosmological models particularly difficult, since the standard technique is to run computationally expensive n-body simulations that calibrate matter power spectrum emulators and fitting functions. This is necessary since only a limited amount of information is captured by Fourier modes for which linear theory accurately describes gravitational collapse.  For the case of coupled dark energy, there have been several attempts to simulate the nonlinear matter power spectrum \cite{Saracco:2009df,Casas:2015qpa}, but these have been developed only for a limited set of cosmological parameters and choices of the quintessence potential. Emulators and fitting functions must be accurate on the entire volume of parameter space with non-negligible posterior probability. This calibration requires many simulations for each potential and functional form of the dark sector coupling. For example, the coyote $\Lambda$CDM emulator was calibrated against 37 state-of-the-art n-body simulations in addition to dozens of predictions from renormalized cosmological perturbation theory  \cite{Heitmann:2013bra,Lawrence:2017ost}. Failing to enforce similar accuracy requirements for coupled dark energy models would result in constraints on the coupling parameter being limited by systematic uncertainties.

The difficulty of developing power spectrum emulators for every possible scalar-field potential and coupling can be alleviated by the use of semi-analytical techniques. However, these methods are limited to quasi-linear Fourier modes $k \lesssim 0.8 \, \text{h/Mpc}$. Motivated by this limitation, we would like to address the question of whether DES and LSST lensing measurements on scales that are accessible to semi-analytical techniques can significantly tighten constraints from Planck data on the coupling of dark energy to dark matter. 

For definiteness, we analyze the case of a conformal coupling which may arise naturally from higher dimensional theories with branes, such as the Randall Sundrum I model \cite{Randall:1999ee} and in Brans-Dicke theories after a conformal transformation \cite{Amendola:1999qq,Wetterich:1994bg}. Indeed, having a field-theoretical description in mind can be useful in a number of circumstances, particularly when understanding the limits of applicability of the model~\cite{CarrilloGonzalez:2017cll}. Although we specialize to a particular coupling and quintessence potential, which have been the subject of past investigations, we believe that our results apply to more general parameterizations as argued in the following paragraphs.

A simple way to model dark energy coupled to dark matter is to treat both components as perfect fluids. The energy-momentum tensors of dark matter and dark energy, instead of being conserved independently, satisfy
\begin{equation}
\nabla_\mu T^{\mu\nu}_\text{cdm}=-\nabla_\mu T^{\mu\nu}_\text{de}=Q^\nu=\xi H u^\nu \rho_\text{cdm/de} \ ,
\label{coupledemtensors}
\end{equation}
where $\rho_\text{cdm/de}$ either stands for $\rho_\text{cdm}$, the dark matter density, or $\rho_\text{de}$, the dark energy density. Here $H$ is the Hubble parameter; and $\xi$ is usually taken to be a constant, although the more fundamental field theory models we describe below can give rise to a non-constant $\xi$. Within the fluid description, the above equations can be written as
\begin{eqnarray}
\frac{d\rho_\text{cdm}}{dt}+3H\rho_\text{cdm}=Q \label{Qcdm}\\ 
\frac{d\rho_\text{de}}{dt}+3H(1+w_\text{de})\rho_{de}=-Q \label{Qde}\ ,
\end{eqnarray}
where $Q=\xi H \rho_\text{cdm/de}$ and $w_{de}$ is the equation of state parameter for the dark energy component. Thus, if $Q>0$, energy is transferred from dark energy to dark matter, and if $Q<0$, the situation is reversed. 

While different couplings can give rise to different expressions for $\xi$, their effects will often be somewhat similar. This can be seen by writing the equations of motion for the dark sector as
\begin{equation}
\begin{split}
\frac{d\rho_\text{cdm}}{dt}+3H\rho_\text{cdm}\big(1+w_\text{cdm}^\text{eff}\big)&=0 \ ,\\
\frac{d\rho_\text{de}}{dt}+3H\rho_\text{de}\big(1+w_\text{de}^\text{eff}\big)&=0 \ .
\end{split}
\end{equation}
Assuming, for example, that $Q\propto\rho_{\text{cdm}}$, we have that the effective equations of state are
\begin{equation}
\begin{split}
w_\text{cdm}^\text{eff}&=-\frac{\xi H}{3} \ , \\
w_\text{de}^\text{eff}&=w_\text{de}+\frac{\xi H}{3}\frac{\rho_\text{cdm}}{\rho_\text{de}} \ .
\end{split}
\end{equation} 
The definitions for the effective equations of state for the case $Q\propto\rho_{\text{de}}$ follow similarly, although we do not expect our results to hold for these models. On the other hand, considering different quintessence potentials for the dark energy field can change the total equation of state, both today and during the accelerated epoch, but we do not expect these differences to change the main conclusions of this paper.

The outline of this paper is the following. In section \ref{coupled}, we introduce the coupled dark energy models under consideration. In section \ref{constraints}, we show the constraints from current CMB data and the lensing forecast; for both cases, we first explain the basic setup used to analyze the data and then state our results. We show that, by combining the CMB and lensing data, it is possible to probe the dark sector coupling to a few percent using only mildly non-linear scales. Lastly, we discuss our results in section \ref{discussion}.

\section{Coupled dark energy} \label{coupled}
We will focus, for definiteness, on models in which there is a conformal coupling between the dark energy and the dark matter; furthermore, we assume no direct coupling between the dark sector and the standard model. The action reads
\begin{equation}
S=\int\ud^4x\sqrt{-g}\left[\frac{M_\text{pl}^2}{2}R-\frac{1}{2}\left(\nabla\phi\right)^2-V(\phi)\right] \\
+S\left[e^{\alpha(\phi)/2} g_{\mu\nu},\rho_{c}\right]+\sum_jS_j\left[g_{\mu\nu},\psi_j\right] \ , \label{inter}
\end{equation}
where $\phi$ is the dark energy (or quintessence) field, $\rho_{c}$ is the dark matter energy density, and $\psi_j$ represent the standard model fields.  By minimally coupling the standard model fields to the Einstein-frame metric, we ensure that the model satisfies equivalence principle constraints from, for example, solar system tests of gravity.  Assuming a flat FRW metric $ds^2=g_{\mu\nu}dx^{\mu}dx^{\nu} = a(\eta)^2(-d\eta^2+ d{\bf x}^2)$, with conformal time $\eta$, the equations of motion involving the dark components become
\begin{eqnarray}
&3M_\text{pl}^{2}\displaystyle \mathcal{H}^{2}=\frac{1}{2}\dot{\phi}^{2}+a^2V(\phi)+e^{\alpha(\phi)}a^2\rho_{c} + a^2\sum_i \rho_i \ , \\
&\ddot{\phi}+2\mathcal{H}\dot{\phi}+a^2V'(\phi)=-a^2\alpha'(\phi)e^{\alpha(\phi)}\rho_{c} \ , & \\
&\dot{\bar\rho}_{c}+3\mathcal{H}\bar\rho_{c}=\alpha'(\phi)\dot\phi\bar\rho_{c} \ , & 
\label{cdmeq}
\end{eqnarray}
where a prime denotes a derivative with respect to $\phi$, an overdot denotes a derivative with respect to the conformal time, $\mathcal{H}\equiv {\dot a}/a$ is the conformal Hubble parameter, and the subscript $i$ denotes that the sum is over all standard model particles, including neutrinos, which can also have a nontrivial effect on the interpretation of our results. Finally,
\begin{equation}
\bar\rho_{\text{c}}\equiv e^{\alpha(\phi)}\rho_{\text{c}},
\end{equation} 
is the observed dark matter density, which is not conserved.

In previous studies, the existence of attractors has been shown for different quintessence potentials \cite{Amendola:1999er, Bean:2008ac}. Here, we will focus on the case of an exponential potential given by
\begin{equation}
\label{eqn:exponentialpotential}
V(\phi)=V_0 e^{-\lambda\phi/M_{\text{pl}}} \ ,
\end{equation}
where $\lambda$ is a dimensionless positive constant and $V_0$ is a positive constant with units of $[{\rm mass}]^4$. We will also specialize to coupling functions of the form
\begin{equation}
\label{eqn:coupling}
\alpha(\phi)= -C \sqrt{\frac{2}{3}} \frac{\phi}{M_{\text{pl}}} \ ,
\end{equation}
where the factor of $\sqrt{2/3}$ is introduced for convenience and where $C$ is a dimensionless constant. In this work, for simplicity, we will restrict ourselves to a parameter space that excludes the case where the scalar field and radiation are both comparably dominant before radiation-matter equality, this requires $|C|<\sqrt{3}\big/2$. We will also require the existence of a matter dominated phase and a late-time accelerator attractor with an effective equation of state close to $\Lambda$CDM. There are two attractors that can give rise to an accelerated epoch, \cite{Bean:2008ac,Amendola:1999er}. In order to ensure that we reach the required accelerated epoch we require $0<\lambda<2 C$. Therefore, we limit the analysis to $0<\lambda<\sqrt{3}$ and $|C|<\sqrt{3}/2$.

To simplify comparisons with wCDM models, we define effective densities in such a way that what we refer to as dark matter redshifts exactly as in the standard CDM case:
\begin{equation}
\begin{split}
\rho_{\text{DE}}&\equiv\frac{1}{2}\dot{\phi}^{2}+V(\phi)+(e^{\alpha(\phi)}-1)\rho_{\text{c}} \ , \\ 
\rho_\text{\text{DM}}&\equiv\rho_\text{c} \ .
\end{split}
\end{equation}
With these definitions, if we assumed that the universe contained only non-interacting dark components, we would infer
\begin{eqnarray}
\label{eqn:effective_background}
w_{\text{DE}} &=& \frac{\frac{1}{2}\dot{\phi}^{2}-V(\phi)}{\rho_{\text{DE}}} \ , \\
\Omega_{\text{DE}} &=& \frac{\rho_{\text{DE}}}{\rho_{\text{DE}} + \rho_\text{\text{DM}} + a^2\sum_i \rho_i} \ , \\
\Omega_{\text{DM}} &=& \frac{\rho_{\text{DM}}}{\rho_{\text{DE}} + \rho_\text{\text{DM}} + a^2\sum_i \rho_i} \ . 
\end{eqnarray}
The results in this paper are obtained, of course, by considering the perturbed versions of the equations of motion of the above model, these equations are found in Appendix \ref{ps}. We present a detailed computation of the adiabatic and dark sector isocurvature modes in Appendix \ref{shp} and note here that, when solving the perturbed equations, we assume adiabatic initial conditions.

The models of coupled dark energy under consideration have been extensively studied; see for example \cite{Amendola:1999er,Bean:2008ac,Koivisto:2005nr,Amendola:2003eq,Amendola:2003wa,Amendola:2002bs,Bean:2007nx,Bean:2007ny, LaVacca:2009yp,Pettorino:2008ez}. A general feature of these models is the dilution of the dark matter energy density at a rate faster than $a^{-3}$. In addition, the interaction between the dark sectors increases the distance to the last-scattering surface, which in turn shifts the position of the acoustic peaks of the CMB temperature power spectrum to larger multipoles. Compared to the non-interacting case, the presence of the coupling leads to a later matter-radiation equality and smaller observed dark matter density fraction: $\bar{\Omega}_m = \bar{\rho}_c/\rho_\text{critical}$. In the following section, we will argue that these and other features mean that the coupling can mimic the effect of changing the spectral index on the linear matter power spectra. 


\section{Constraining the coupling strength} \label{constraints}

Having defined the class of models under consideration, in this section, we show how CMB temperature and polarization data in combination with lensing and clustering of galaxies measurements from DES and LSST can tighten the current state-of-the-art constraints on the coupling $C$ in~(\ref{eqn:coupling}). We also examine the potential impact of including low redshift information from baryon acoustic oscillations (BAO) and type IA supernova data, and show that these do not significantly improve the upper limits on the coupling (see figure~\ref{fig:Plot1}). We understand this to result from the high level of degeneracies between the parameters involved.

\subsection{Constraints from current data}

\subsubsection{Basic Setup}
To understand current constraints we need to focus on specific datasets and likelihood functions. For the CMB data, we adopt the Planck low-$\ell$\footnote{Low-TEB likelihood, $\ell < 30$ for the TT, TE and EE spectra.} and high-$\ell$\footnote{Plik likelihood, $30 < \ell < 2508$ for the TT spectrum, and  $30 < \ell < 1996$ for the TE and EE spectra.} likelihoods \cite{Ade:2015xua,2016A&A...594A..11P}. For the BAO data, we adopt the DR11, 6DF and MGS datasets~\cite{2014MNRAS.441...24A,2011MNRAS.416.3017B,2015MNRAS.449..835R}. Finally,  for the type IA supernova data, we adopt the JLA compilation~\cite{2014A&A...568A..22B}. The implementations of both the BAO and the JLA likelihoods were already included in the November 2016  version of the CosmoMC~\cite{COSMOMC} software.

To compare our model to these datasets, we first modified the CAMB code (November 2016 version)~\cite{CAMB, 2012JCAP...04..027H, 2000ApJ...538..473L} to evaluate the temperature and polarization power spectra predicted by coupled dark energy models. We then estimated the posterior via Monte Carlo Markov Chains that were implemented using the CosmoMC code~\cite{2013PhRvD..87j3529L, 2002PhRvD..66j3511L}. The numerical routines used to implement our modifications to CAMB were provided by the NAG Fortran library~\cite{NAGFORTRAN}.

Our baseline cosmology contains nine parameters $\{\Omega_{\text{DM}} h^2, \Omega_b h^2, \theta_{\text{A}}, n_s,  A_s, \tau, \sum m_\nu, C, \lambda\}$.
Here, $h$ is defined in terms of the Hubble constant via $H_0\equiv 100h\rm{km/s/Mpc}$, $\Omega_{\text{DM}} h^2$ and $\Omega_b h^2$ are the effective cold dark matter and baryon energy densities respectively, $\theta_A$ is the angular acoustic scale at recombination, $n_s$ and $A_s$ are the inflationary spectral index and power spectrum amplitude respectively, $\tau$ is the optical depth at reionization, $\sum m_\nu$ is the sum of neutrino masses, $C$ is the dark energy coupling defined in~(\ref{eqn:coupling}), and $\lambda$ is the potential parameter defined in~(\ref{eqn:exponentialpotential}). As mentioned earlier, we are assuming spatial flatness. 

We include the sum of neutrino masses as a free parameter since there are well-known degeneracies between certain effects of massive neutrinos and those of particular models of modified gravity~\cite{2014MNRAS.440...75B,2017JCAP...02..043B}. This allows us to explore the question of whether such degeneracies are also present in our models, and if so, the extent to which they affect our ability to constrain the coupling parameter. Given that we are still not at the level of precision at which differences between the normal, degenerate, and inverse hierarchies are statistically significant and prior-independent~\cite{2017arXiv170303425S,2017JCAP...02..043B,2017arXiv170310829C}, we assume the degenerate hierarchy for simplicity. Finally, since we have assumed that neutrinos are not coupled directly to the scalar field (or the dark matter), neutrinos and cold dark matter behave differently at all scales. This is a significant difference in comparison to $\nu$CDM cosmologies, where effects on the matter power spectrum are indistinguishable above the neutrino free-streaming scale~\cite{1998PhRvL..80.5255H,2014PhRvD..90h3530L}.

\subsubsection{Results}

\begin{figure*}[!h]
	\includegraphics[width=0.75\linewidth]{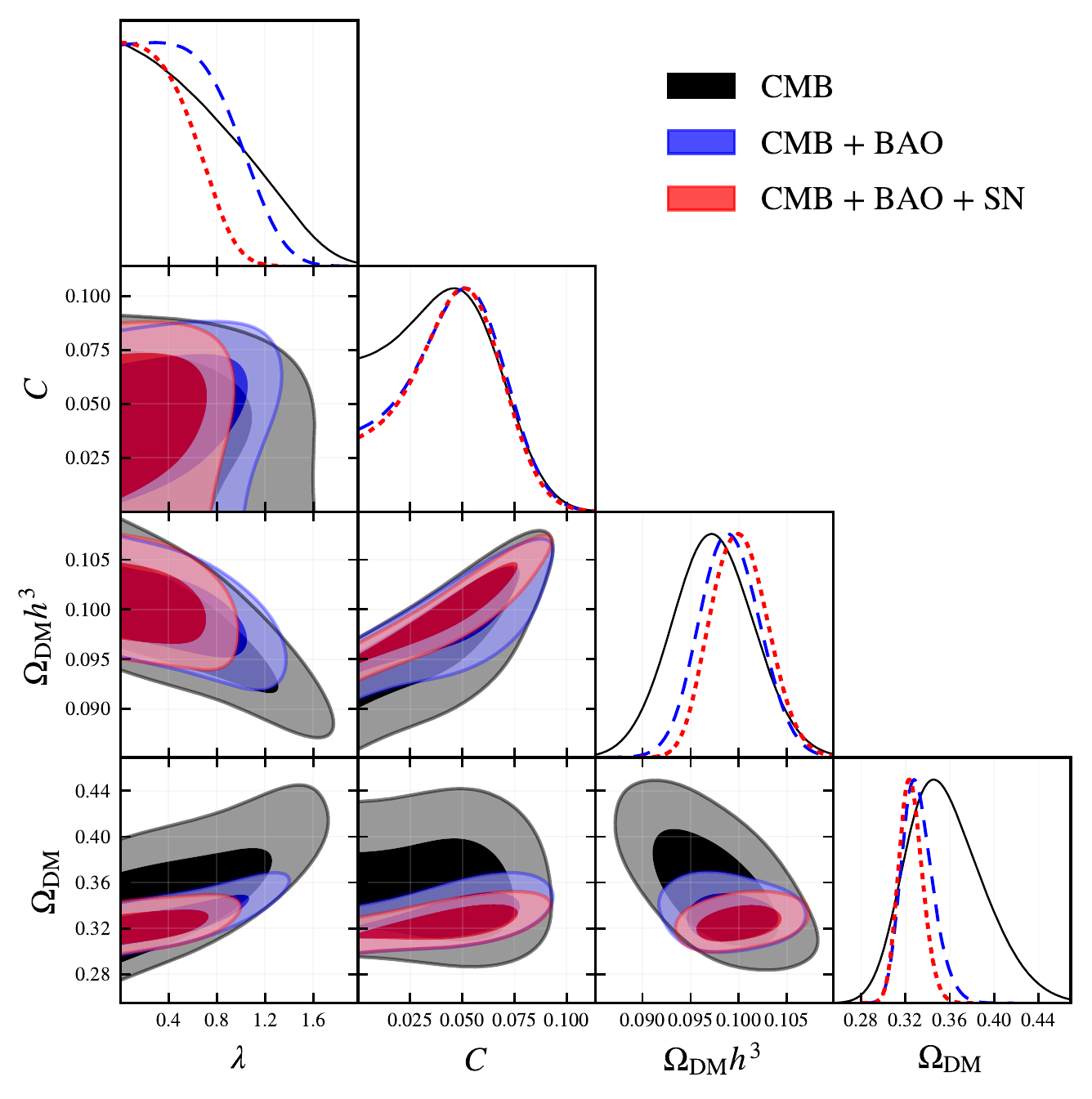}
	\caption{The posterior distribution for the cosmological parameters in the context of coupled dark energy models. The CMB data consists of both temperature and polarization data, but excludes lensing reconstruction. Similar to the results shown in~\cite{Ade:2015rim}, the combination of the CMB, BAO and type IA supernova low redshift data does not tighten the upper limits on $C$ in comparison with the constraints coming from the CMB alone.}
	\label{fig:Plot1} 
\end{figure*}

\begin{figure*}[!h]
	\includegraphics[width=0.7\linewidth]{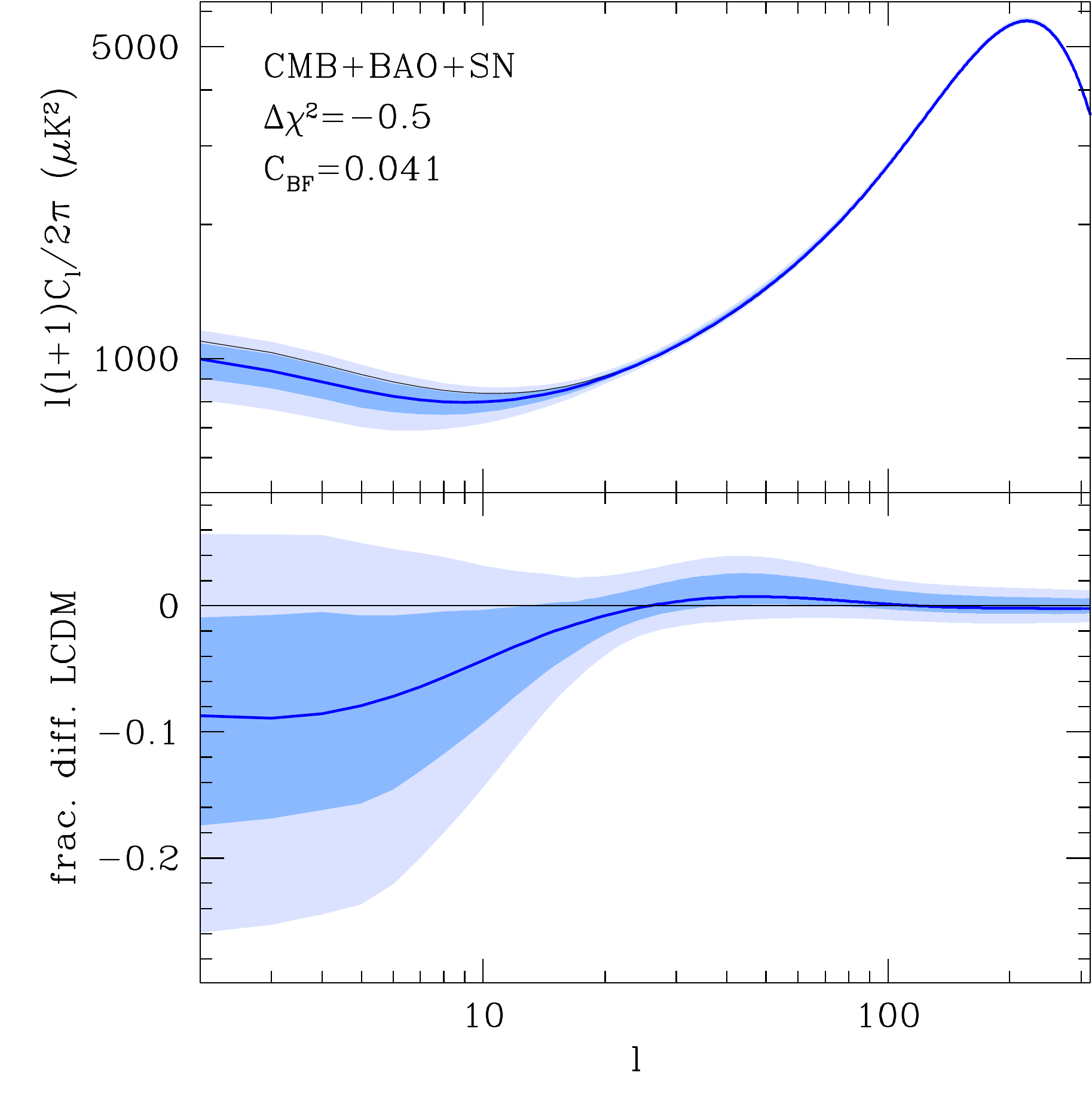}
	\caption{The best-fit model (blue solid line) that has $C=0.041$ and the posterior distribution (one and two sigma contours are the dark and light blue bands respectively) for the CMB temperature power spectrum predicted by coupled dark energy models. We have also included the best-fit $\Lambda$CDM model (solid black line) to show the lack of power at large scales, which is a well known feature of the Planck data and which explains the weak preference for positive couplings. The difference in $\chi^2 \equiv - 2\ln \mathcal{L}$ between the best-fit models is, however, not large enough to claim any detection, even at the one sigma level.} 
	\label{fig:Plot2} 
\end{figure*}

In figure~\ref{fig:Plot1}, we show the results obtained by using only the CMB data,  then combining this first with BAO measurements, and finally including supernovae data. We find that temperature and polarization data from Planck can rule out couplings $C \gtrsim 0.1$, similar to the constraints obtained in the Planck collaboration paper~\cite{Ade:2015rim} that analyzed coupled scalar field models with an inverse power-law potential. We can observe a weak preference for a non-zero coupling that we infer is driven by the preference for low power in the Planck temperature spectrum at large scales, which can be seen in figure~\ref{fig:Plot2}. Finally, there is a correlation between the coupling and $\Omega_{\text{DM}} h^3$, which is due to the effects of both parameters on the angular size of the acoustic peaks at recombination.

\begin{figure*}[!h]
	\includegraphics[width=0.75\linewidth]{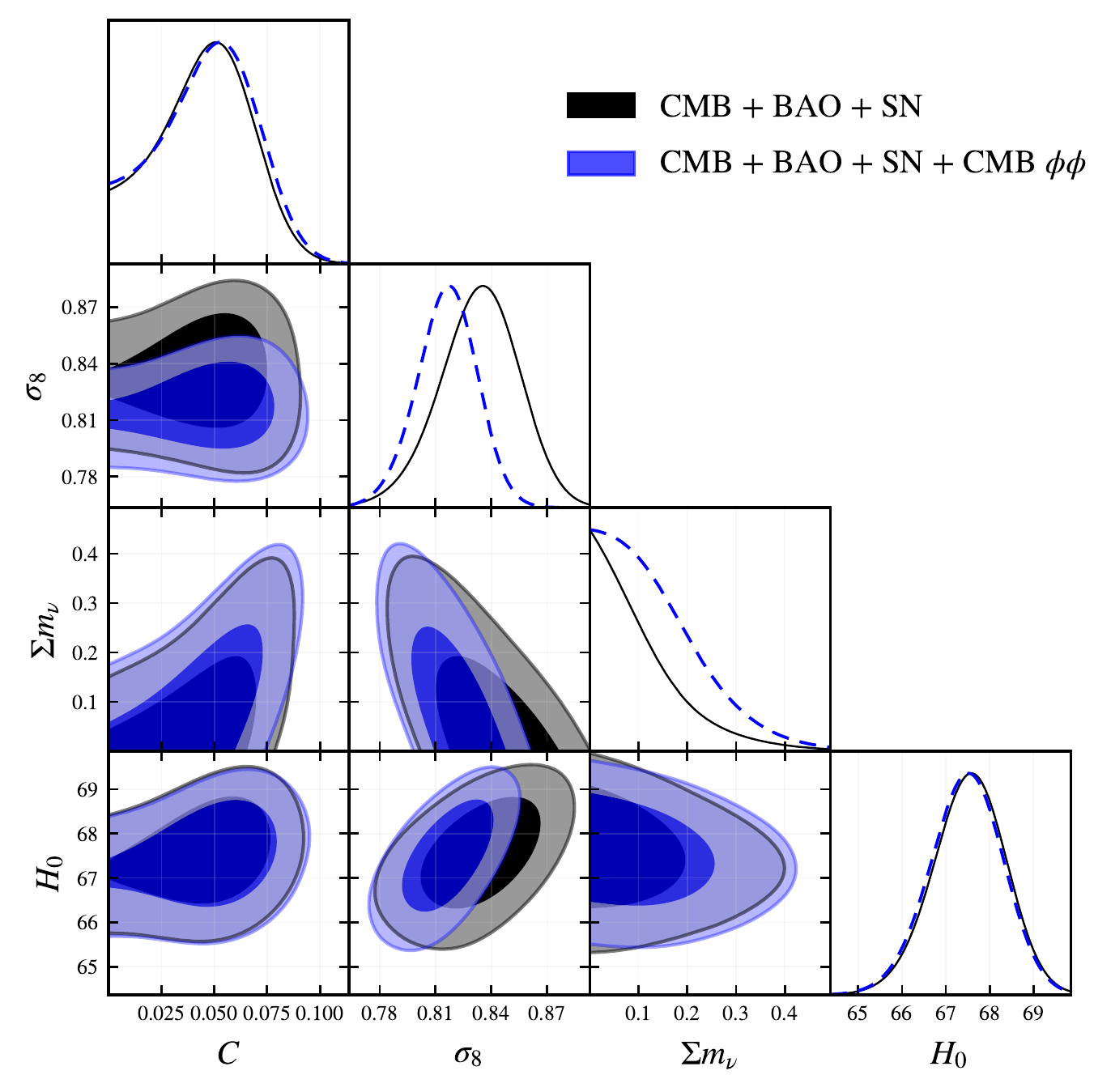}
	\caption{The posterior distribution for cosmological parameters in the context of coupled dark energy models. Here $H_0$ and $\sum m_\nu$ are shown in km/s/Mpc and eV respectively. The CMB data consists of both temperature and polarization data, but excludes lensing reconstruction in the black contours. Note that the correlation between $H_0$ and $\sigma_8$ is the opposite sign to that required to alleviate the tensions between low redshift probes and the CMB, i.e. higher $H_0$ correlates with higher $\sigma_8$. Therefore, had we included local $H_0$ measurements, the posterior for non-zero neutrino mass would have been boosted towards larger masses, which leads to a preference for lower $\sigma_8$. Massive neutrinos, however, are not the perfect solution to this, because they also tend to reduce $H_0$ in $\nu\Lambda$CDM, although this correlation is weaker in our models. Indeed, the preference for lower $\sigma_8$\footnote{The lensing potential is slightly incompatible with the amount of lensing extracted from the smoothing of the peaks in the temperature power spectrum and that implies lower $\Omega_{\text{DM}}^{1/2}\sigma_8$ \cite{Obied:2017tpd}.}  broadens the neutrino mass distribution in $C>0$ models by almost a factor of two in comparison with uncoupled models - see the correlation plot between $\sum m_\nu$ and C located in left middle panel.} 
\label{fig:Plot3} 
\end{figure*}

The addition of BAO and type IA supernova data provides extra information on $\Omega_{\text{DM}}$, allowing for an improvement of the upper limit on $\lambda$ due to the correlation between $\Omega_{\text{DM}}$ and $\lambda$ shown in the bottom-left panel on figure~\ref{fig:Plot1} (low $\Omega_{\text{DM}}$ is not compatible with high values of $\lambda$). Notice that the upper limit on the coupling is mainly unaffected by the addition of the low-redshift data. The Planck collaboration has shown similar results (see figure 21 on \cite{Ade:2015rim}) in the context of an inverse power-law potential. On the other hand, the preference for a positive nonzero coupling does get slightly stronger, due to the preference for lower $\Omega_{\text{DM}}$. Indeed, lower values of $\Omega_{\text{DM}}$ suggest higher $\Omega_{\text{DM}} h^3$, thus slightly disfavoring $C\lesssim0.05$. Lower $\Omega_{\text{DM}}$ also excludes models with large neutrino masses. Nevertheless, comparison of the maximum likelihood model against $\Lambda$CDM shows small differences in $\chi^2$. All the statements listed above also hold in chains with fixed neutrino masses ($\sum m_\nu=0.06$eV).

The correlation between the sum of neutrino masses and the dark energy coupling emerges when data from the CMB and BAO are combined. Indeed, higher coupling values can relax the upper limits on the neutrino masses by a factor of two. The main reason for this lies in the fact that the correlation between $\sigma_8$ and $H_0$ is opposite to that required to simultaneously alleviate the existing tensions between the CMB data and that from low redshift probes, including direct measurements of $H_0$. The comparison shown in figure~\ref{fig:Plot3} demonstrates that higher neutrino masses imply lower $\sigma_8$, which alleviates the tension.

The results discussed in this section broadly reproduce what is known about the effects of coupled models on cosmological parameters. We now turn to the question of how these constraints might be improved by including lensing data from current and upcoming cosmological surveys.

\subsection{Lensing forecast}

\subsubsection{Basic Setup}

Our central interest is in what we might learn from the DES and LSST surveys. The primary goal is to understand whether lensing can improve our constraints on the coupling between dark energy and dark matter without taking into account scales that are inaccessible to semi-analytical perturbation methods, such as renormalized perturbation theory~\cite{2006PhRvD..73f3519C} and effective field theories of large scale structure~\cite{2012JHEP...09..082C,2014JCAP...07..057C}.

The detailed settings for the DES and LSST surveys are listed in tables~\ref{tab:params_des} and ~\ref{tab:params_lsst} respectively. The large-scale structure observables we include in our forecast are the cosmic shear, galaxy lensing, and galaxy clustering \cite{2004PhRvD..70d3009H}. In all chains, the fiducial model adopted for the lensing forecasts is the flat $\Lambda$CDM one, with parameters shown in table~\ref{table:fiducial_parameters}. The covariance matrices for DES and LSST, as well the data vector evaluated in each chain step, were calculated using a modified version of CosmoLike~\cite{2016arXiv160105779K} adapted to receive the matter power spectrum and the background distances from our modified CAMB code.

The posterior was sampled with the multimodal nested sampling algorithm MultiNest \cite{2009MNRAS.398.1601F} instead of the dragging method that is the default algorithm in the CosmoMC code~\cite{2013PhRvD..87j3529L}. This implementation was developed in CosmoSiS~\cite{2015A&C....12...45Z}, which provides a robust framework to exchange parameters between Multinest, CAMB, CosmoLike, and the likelihoods.  These changes were not motivated in any way by the shape of the lensing likelihood, which is a high-dimensional multivariate normal distribution that can be adequately sampled using either methods. Rather, the choice was mainly due to the complexity of the pipeline and the different computer languages involved in the development of the various modules\footnote{Fortran (CAMB), C (CosmoLike) and Python (Likelihoods).}. In such a setup, the CosmoSiS framework is more suitable for the transferring of data between the modules\footnote{CosmoSiS implements the Multinest algorithm but does not offer an implementation of the CosmoMC sampler.}

\begin{table}[!h]
	\caption{Settings of the DES forecast}
		\def\arraystretch{0.60}
	\begin{center}
		\begin{tabular*}{0.8\textwidth}{@{\extracolsep{\fill}}| c c c |}
			\hline
			\,\, Parameter & Fiducial Value & Prior \\  
			\hline
			\hline
			\multicolumn{3}{|c|}{Basic Settings} \\
			\hline
			\,\, Survey Area & 5,000 deg$^2$ & fixed\\
			Source Ellipticity Dispersion $\sigma_\epsilon$ &0.37& fixed\\
			Project Source Density $n_\text{gal}$ &10.0 galaxies/arcmin$^2$& fixed\\
			Project Lens Density $n_\text{gal}^\text{lens}$ &0.15 galaxies/arcmin$^2$& fixed\\
			\hline
			\hline
			\multicolumn{3}{|c|}{Galaxy Bias} \\
			\hline
			$b_\text{g}^1$ & 1.35  & flat (0.1, 10.0) \\
			$b_\text{g}^2$ & 1.5  &flat (0.1, 10.0) \\
			$b_\text{g}^3$ & 1.65 & flat (0.1, 10.0) \\
			\hline
			\hline
			\multicolumn{3}{|c|}{Lens Photo-z} \\
			\hline
			$\Delta_\text{z}^i$  (photo-z bias) & 0 &  flat (-0.0005, 0.0005)\\ 
			$\sigma_\text{z} $ & 0.01 & fixed \\
			\hline
			\hline
			\multicolumn{3}{|c|}{Source Photo-z} \\
			\hline
			$\Delta_\text{z}^i $  (photo-z bias) & 0 & flat (-0.1, 0.1) \\
			$\sigma_\text{z}$ & 0.05 & fixed \\
			\hline
			\hline
			\multicolumn{3}{|c|}{Shear Calibration} \\
			\hline
			$m_i$ (multiplicative bias)  & 0 & flat (-0.01, 0.01)\\
			\hline
		\end{tabular*}
	\end{center}
	\label{tab:params_des}
\end{table}

\begin{table}[!h]
	\caption{Settings of the LSST forecast}
		\def\arraystretch{0.60}
	\begin{center}
		\begin{tabular*}{0.8\textwidth}{@{\extracolsep{\fill}}| c c c |}
			\hline
			Parameter &   Fiducial Value   & Prior \\  
			\hline 
			\hline
			\multicolumn{3}{|c|}{Basic Settings} \\
			\hline
			Survey Area & 18,000 deg$^2$ & fixed\\
			Source Ellipticity Dispersion $\sigma_\epsilon$ &0.37& fixed\\
			Project Source Density $n_\text{gal}$ &26.0 galaxies/arcmin$^2$& fixed\\
			Project Lens Density $n_\text{gal}^\text{lens}$ &0.25 galaxies/arcmin$^2$& fixed\\
			\hline
			\hline
			\multicolumn{3}{|c|}{Galaxy Bias} \\
			\hline
			$b_\text{g}^1$ & 1.35  & flat (0.1, 10.0) \\
			$b_\text{g}^2$ & 1.5  &flat (0.1, 10.0) \\
			$b_\text{g}^3$ & 1.65 & flat (0.1, 10.0) \\
			$b_\text{g}^4$ & 1.8 & flat (0.1, 10.0) \\
			\hline
			\hline
			\multicolumn{3}{|c|}{Lens Photo-z} \\
			\hline
			$\Delta_\text{z}^i $  (photo-z bias) & 0 & fixed \\
			$\sigma_\text{z} $ & 0.01 & fixed \\
			\hline
			\hline
			\multicolumn{3}{|c|}{Source Photo-z} \\
			\hline
			$\Delta_\text{z}^i $  (photo-z bias) & 0 & fixed \\
			$\sigma_\text{z}$ & 0.05 & fixed \\
			\hline
			\hline
			\multicolumn{3}{|c|}{Shear Calibration} \\
			\hline
			$m_i $ (multiplicative bias) & 0 & flat (-0.005, 0.005)\\
			\hline
		\end{tabular*}
	\end{center}
	\label{tab:params_lsst}
\end{table}

\begin{table}[!h] 
	\caption {Fiducial $\Lambda$CDM parameters used for both the DES and LSST forecasts.} 
	\def\arraystretch{0.60}
	\begin{center}
	\begin{tabular}{| c | c | }
		\hline 
		$10\Omega_{\text{DM}} h^2$ & $1.199 $ \\
		$ 100\Omega_b h^2$ & $2.222$ \\ 
		$10 n_s$ & $9.652$  \\
		$H_0$ & $67.26$ \\
		$100\tau$ & $7.80$ \\
		$10^9 A_s$ & $2.1985$ \\
		$m_\nu$ & $0$ \\
		\hline
	\end{tabular} 
	\label{table:fiducial_parameters} 
	\end{center}
\end{table} 

Our covariance matrix for the lensing and clustering of galaxies takes into account both gaussian and non-gaussian contributions, as well as the full correlation between these observables (see Appendix A of \cite{2016arXiv160105779K} for details of the implementation). However, we have not modeled nonlinear baryonic effects~\cite{2015MNRAS.454.2451E} or the intrinsic alignment~\cite{2015MNRAS.450.2195S}. We have included photometric redshift uncertainties and shear multiplicative bias systematics (see \cite{2016arXiv160105779K} for implementation details). We note that the $\Lambda$CDM predictions are derived from one of the analysis codes validated in \citep{Krause:2017ekm}.

For the DES forecast, we have assumed a flat gaussian prior between $[-0.01, 0.01]$ for the photometric redshift biases of the source galaxies and a flat prior over the same range for the shear multiplicative biases. We assume a lens galaxy sample consisting of luminous red-sequence galaxies similar to the redMaGiC selection \cite{Rozo:2015mmv}, which are selected to have accurate photometric redshifts with $\Delta_z\lesssim 10^{-3}$. We have checked and confirmed that our constraints on the coupling parameter are robust against pessimistic scenarios in which the maximum allowed values for these systematics are three times larger. Currently, the DES year one analysis assumes priors of a few percentages in these biases.

\subsubsection{The effects of coupling on the linear power spectrum }

Before digging into the forecast results, it is worth understanding qualitatively the changes that a positive coupling induces in the linear power spectrum.  The matter linear power spectrum is defined by~\cite{2008arXiv0802.3688H}
\begin{equation}
\label{eqn:linear_power_spectrum}
\frac{k^3}{2\pi^2} P_k = \frac{4}{25} A_s \left(\frac{G(a) a}{\bar{\Omega}_m}\right)^2 \left(\frac{k}{H_0}\right)^4 \left(\frac{k}{k_{\text{norm}}}\right)^{n_s-1} T^2(k) \ ,
\end{equation}
where $G(a)$ is the growth rate relative to the growth rate during the matter dominated epoch in $\Lambda$CDM models, $T(k)$ is the transfer function, $A_s$ and $n_s$ are the inflationary amplitude and the spectral index respectively, and $\bar{\Omega}_m$ is the observed cold dark matter density fraction.  The $4/25$ pre-factor and the $1\big/\bar{\Omega}_m^2 H_0^4$ dependencies come from the Poisson equation that relates the cold dark matter density to the potential $\Phi$. The origin of the other factors can be understood by first realizing that, in Newtonian gauge, assuming cold dark matter domination, we have
\begin{equation}
\frac{k^3}{2\pi^2} P_k = \frac{4}{9}\frac{a^2 k^4}{\bar{\Omega}_m^2 H_0^2} \left(\frac{k^3}{2\pi^2} P_\Phi\right) \ .
\end{equation}
Furthermore, the gauge transformation relating the Newtonian potential $\Phi$ to the comoving curvature $\mathcal{R}$ in the comoving gauge is $\Phi = 3\mathcal{R}\big/5$, which remains valid in the matter dominated era~\cite{2004astro.ph..2060H}. Finally, the conservation of the comoving curvature outside the horizon implies
\begin{align}
\frac{k^3}{2\pi^2} P_\mathcal{R} = A_s \left(\frac{k}{k_{\text{norm}}}\right)^{n_s-1} \ .
\end{align}
This assumes that pressure fluctuations are negligible beyond the horizon, given that, in the absence of anisotropic stresses \cite{2004astro.ph..2060H}, we have $\dot{\mathcal{R}} \sim -\mathcal{H} \sum_J \delta p_J\big/(\rho_J+p_J)$. The presence of large isocurvature perturbations on super-horizon scales, which can be generated during inflation for example, can spoil the conservation of the comoving curvature (see appendix~\ref{shp} for isocurvature pseudo-initial conditions).

\begin{figure}[!h]
	\includegraphics[scale=0.435]{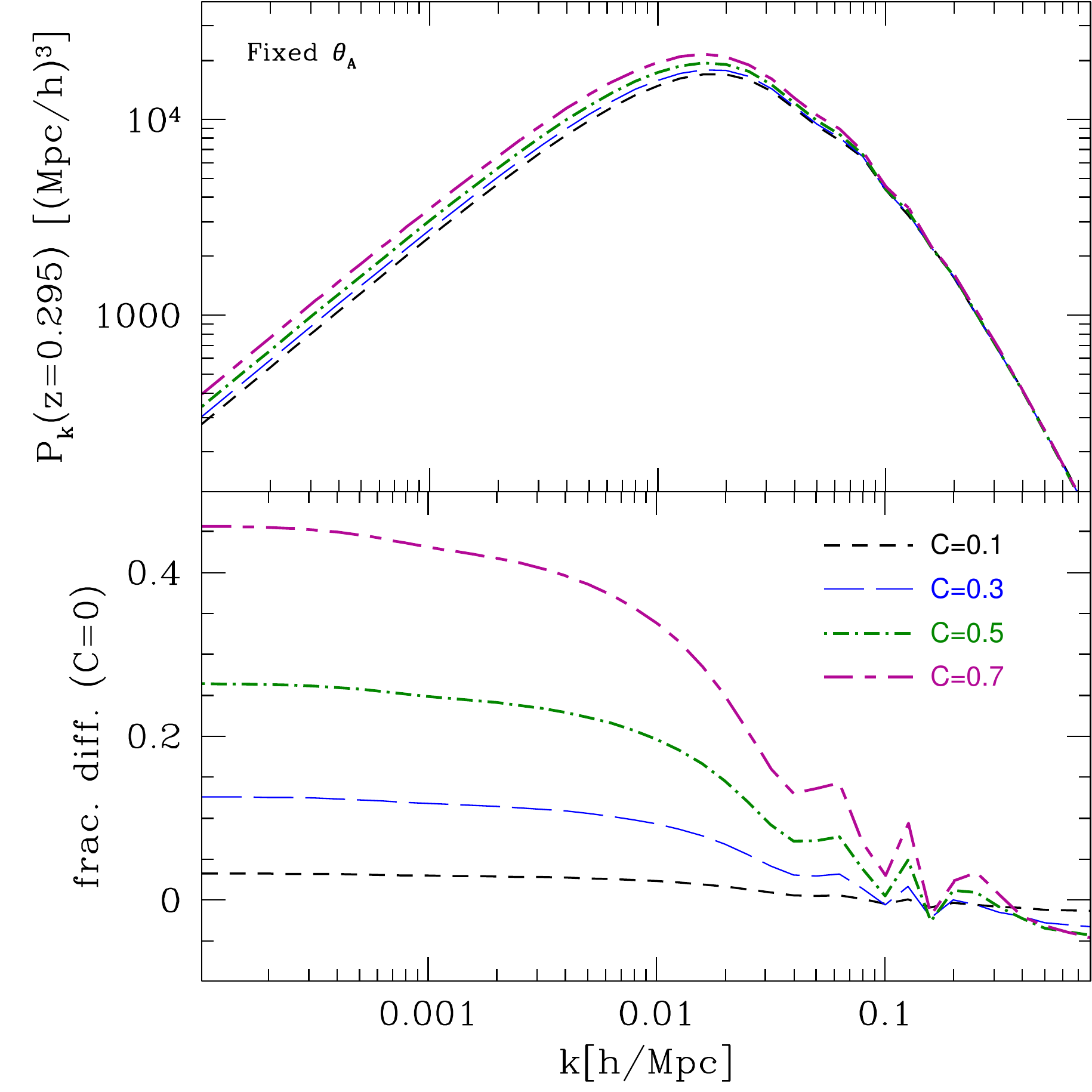}
	\includegraphics[scale=0.435]{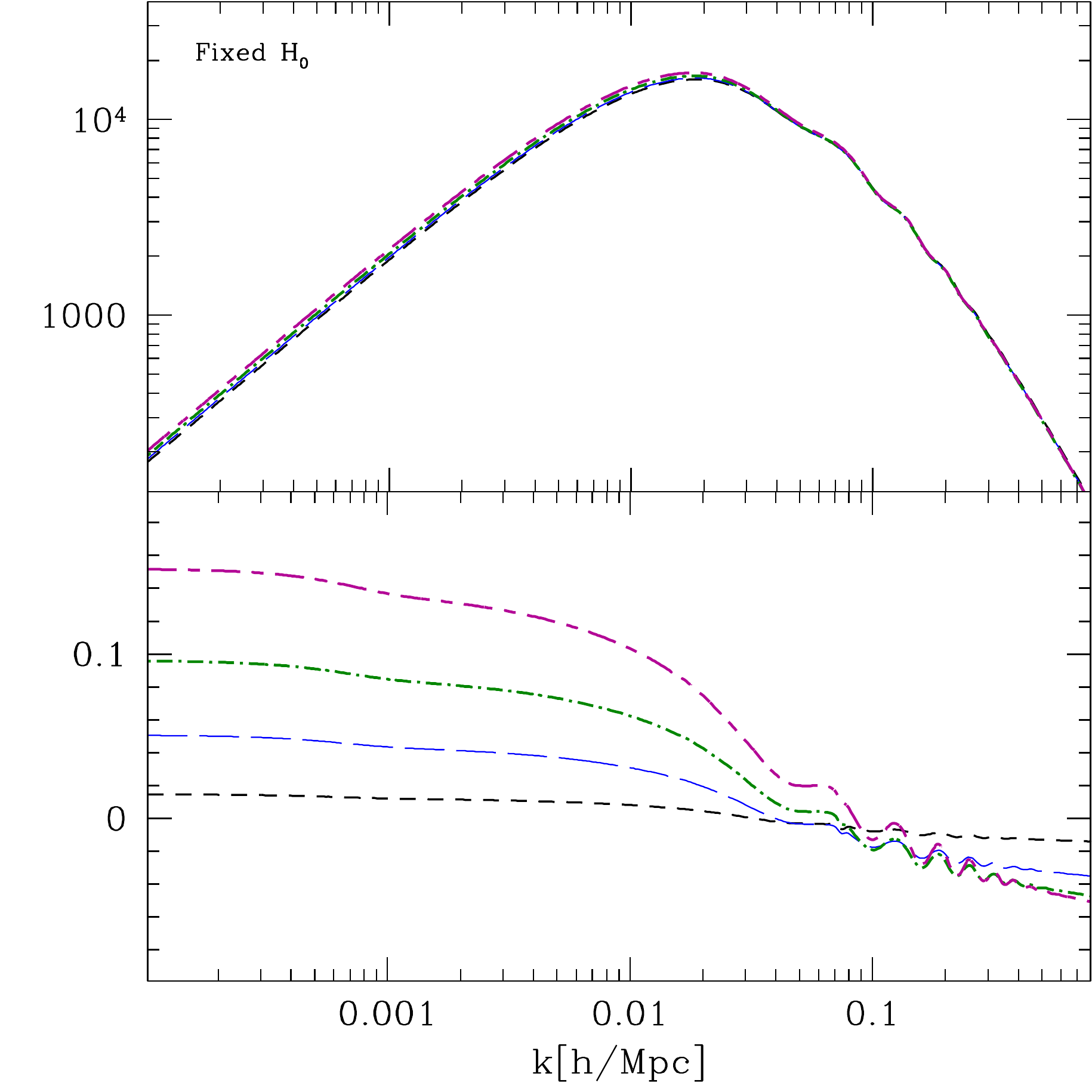}
	\caption{The linear power spectrum for multiple coupled dark energy models with the same $\{\Omega_{\text{DM}} h^2, \Omega_b h^2, n_s,  A_s, \tau, \sum m_\nu, \lambda\}$ cosmological parameters and different values of the coupling $C$. On the right panel, we have fixed $H_0$ since lensing and clustering of galaxies alone does not probe the angular diameter distance to the surface of last scattering. The redshift $z=0.295$ is approximately the median of the first photometric redshift bin for the lenses in the DES survey. For lensing and clustering of galaxies that probes the power spectrum around $0.01 \, \text{h/Mpc} \lesssim k \lesssim 0.5 \, \text{h/Mpc}$, the coupling indeed mimics a change in the inflationary spectral index.}
	\label{fig:Plot4} 
\end{figure}

In figure~\ref{fig:Plot4}, we present the linear matter power spectrum at a redshift that approximately corresponds to the median of the first redshift bin of the lenses in the DES survey. This plot shows that the coupling mimics the effects of changing the inflationary spectral index at the scales probe by DES. There are two underlying reasons for this. Firstly, higher couplings lower $\bar{\Omega}_m$ and this shifts the entire power spectrum upwards\footnote{In $\Lambda$CDM, $\Delta G \sim \frac{1}{2} \Delta\Omega_{\text{DM}}$ around the fiducial value of $\Omega_{\text{DM}}$; the change in the growth factor will also modify the matter power spectrum.}. In comparison to $\Lambda$CDM, coupled dark energy slightly decreases the growth rate, which has the opposite effect. This decrease in the growth rate is expected since $\delta\propto a^{1-4C^2/3}$ grows slower than the uncoupled case, for which $\delta\propto a$ \cite{Amendola:1999er}. The overall effect still shifts the entire power spectrum upwards. Secondly, larger positive couplings also shift matter-radiation equality towards a lower redshift, precisely because there is a lower effective cold dark matter energy density. This moves the location of the power-spectrum turning point to the left, since the universe has a larger horizon size at a lower redshift, and also increases the damping of modes that entered the horizon during the radiation epoch. The latter happens since the shape parameter $\Gamma \equiv \bar{\Omega}_m h$ that controls the transfer function damping~\cite{1999ApJ...511....5E} decreases for positive coupling. The combination of lensing and clustering of galaxies with the CMB data, which fixes $\theta_A$, exacerbates the fractional difference of the linear matter power spectrum with and without coupling. This could be explained by the fact that at fixed $\theta_A$, larger C correlates with larger $H_0$ which in turn induces smaller $\bar{\Omega}_m$ and  $\Gamma$ at fixed $\Omega_{\text{DM}} h^2$.

For negative couplings, the effect on the linear power spectrum is the opposite. Higher $\bar{\Omega}_m$ shifts matter-radiation equality towards higher redshift, which lessens the damping of modes that have entered the horizon during the radiation epoch, and also moves the entire spectrum downwards. Therefore, the linear power spectrum in these models is systematically higher than the $\Lambda$CDM predictions on scales $k \gg 0.01$h/Mpc, and it only crosses the $\Lambda$CDM linear spectrum at $k \ll 0.01$h/Mpc. The lensing signal in this case seems to be degenerate with galaxy bias (see figure 2 of~\cite{Xia:2009zzb}). 

\subsubsection{Results}

\begin{figure*}
	\includegraphics[width=0.7\linewidth]{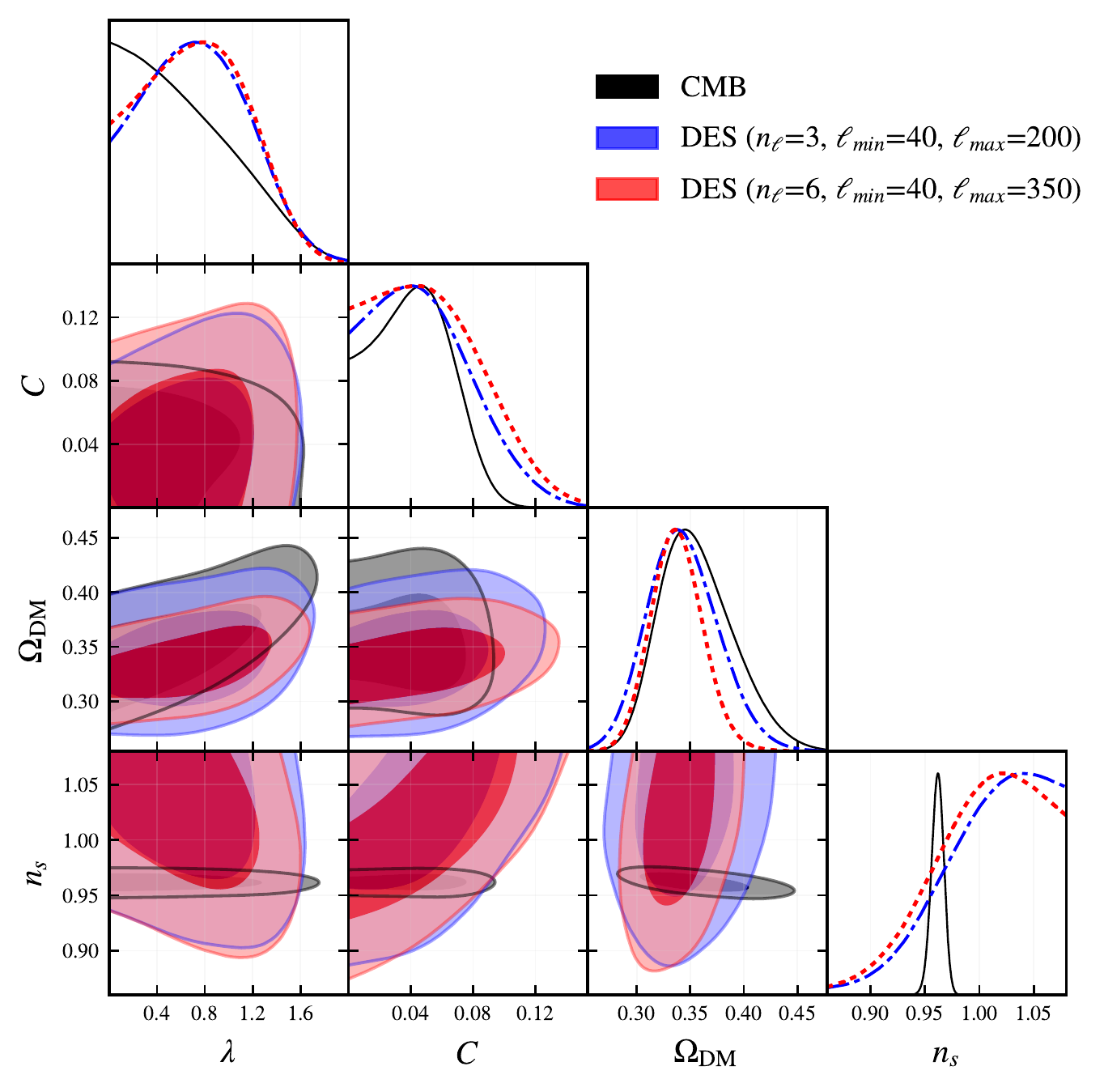}
	\caption{The posterior distribution for cosmological parameters in the context of coupled dark energy models. The CMB data consists of both temperature and polarization data, but excludes lensing reconstruction. In the DES forecasts, $n_\ell$ is the number of bins; $\ell_\text{min}$ and $\ell_\text{max}$ are the minimum and the maximum multipoles. The medians of the redshift bins for the DES lenses are $z=\{0.275,0.425,0.575\}$, and therefore $\ell_\text{max}=350$ corresponds approximately to $k_\text{max} \sim \{0.45,0.3,0.23\} \, \text{h/Mpc}$, within the range accessible to semi-analytical perturbation methods. We have also included a more conservative $\ell_\text{max}=200$ cut, given that $k\sim 0.5 \, \text{h/Mpc}$ requires two-loop calculations in the context of effective field theories to ensure proper modeling~\cite{2014JCAP...07..057C}. Both the $\ell_\text{max}=200$ and $\ell_\text{max}=350$ cases show that the coupling  correlates with changing the inflationary scalar spectral index, and we can use this fact to improve the upper limits on $C$.}
	\label{fig:Plot5} 
\end{figure*}

Given the power of the CMB to constrain the inflationary spectral index to high precision, its combination with galaxy lensing and galaxy clustering is able to improve constraints on the coupling parameter. Indeed, the DES large-scale structure forecasts can, together with a prior of $n_s<1.0$, rule out $C \gtrsim 0.12$, as we can see in figure~\ref{fig:Plot5}. Going from $l_{\max}=200$ to $l_{\max}=350$ reduces the uncertainties in the direction perpendicular to the $n_s-C$ degeneracy in LSST forecast. This is somewhat expected because $l_{\max}=350$ provides sensitivity to a broader range in $k$, and figure~\ref{fig:Plot4} shows that the coupling and the inflationary tilt effects are not perfectly degenerate on the linear power spectrum over many decades in $k$. Therefore, the larger the range probed in $k$ by the data, the better we can constrain the direction perpendicular to the $n_s-C$ degeneracy, as long as the linear power spectrum is a good approximation. On nonlinear scales, however, the matter power spectrum becomes less sensitive to changes in the inflationary spectral index (because of the one-halo term in the halofit approximation), and we therefore see less improvement in the direction perpendicular to the $n_s-C$ degeneracy. 
 
\begin{figure*}
	\includegraphics[width=0.75\linewidth]{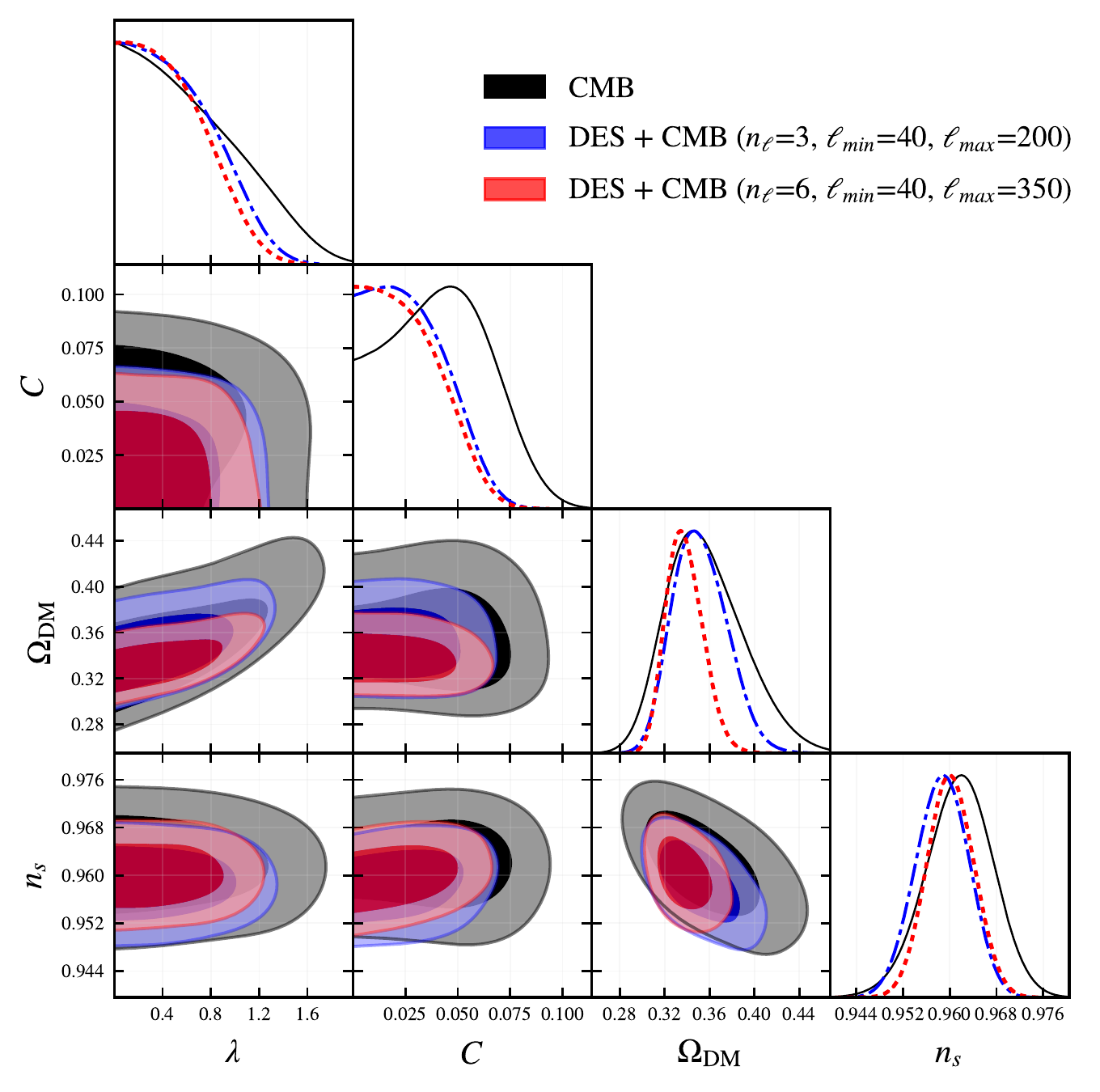}
	\caption{The posterior distribution for cosmological parameters in the context of coupled dark energy models. The CMB data consists of both temperature and polarization data, but excludes lensing reconstruction. This figure shows the final result of combining the CMB and DES lensing, which can improve the upper limits by $25\%$ and also reduce the CMB preference for positive couplings. }
	\label{fig:Plot6} 
\end{figure*}

\begin{figure*}
	\includegraphics[width=0.75\linewidth]{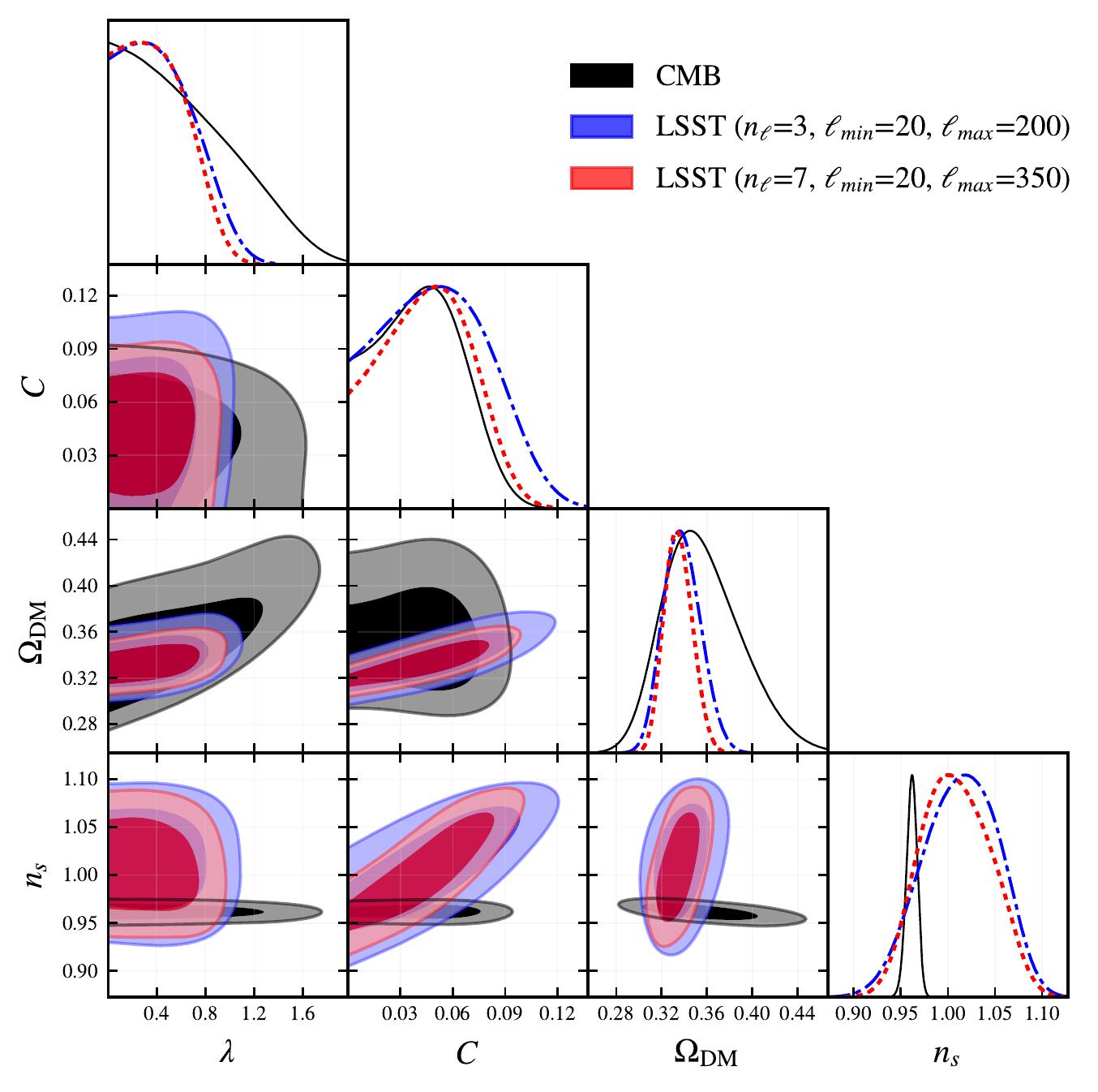}
	\caption{The posterior distribution for cosmological parameters in the context of coupled dark energy models. The CMB data consists of both temperature and polarization data, but excludes lensing reconstruction. The medians of the redshift bins for the LSST lenses are $z=\{0.3,0.5,0.7,0.9\}$, and therefore $\ell_\text{max}=350$ is within the range accessible to semi-analytical perturbation methods. Similarly to the DES forecasts, we have included a more conservative $\ell_\text{max}=200$ cut, given that $k\sim 0.5 \, \text{h/Mpc}$ requires two-loop calculations in the context of effective field theories to ensure proper modeling~\cite{2014JCAP...07..057C}. Again, the $\ell_\text{max}=200$ and $\ell_\text{max}=350$ cases show that the coupling correlates with the inflationary tilt. Here, going from $\ell_\text{max}=200$ to $\ell_\text{max}=350$ does provide advantages that may justify the burden of computing loop corrections in the effective field theory approach.}
	\label{fig:Plot7} 
\end{figure*}

\begin{figure*}
	\includegraphics[width=0.75\linewidth]{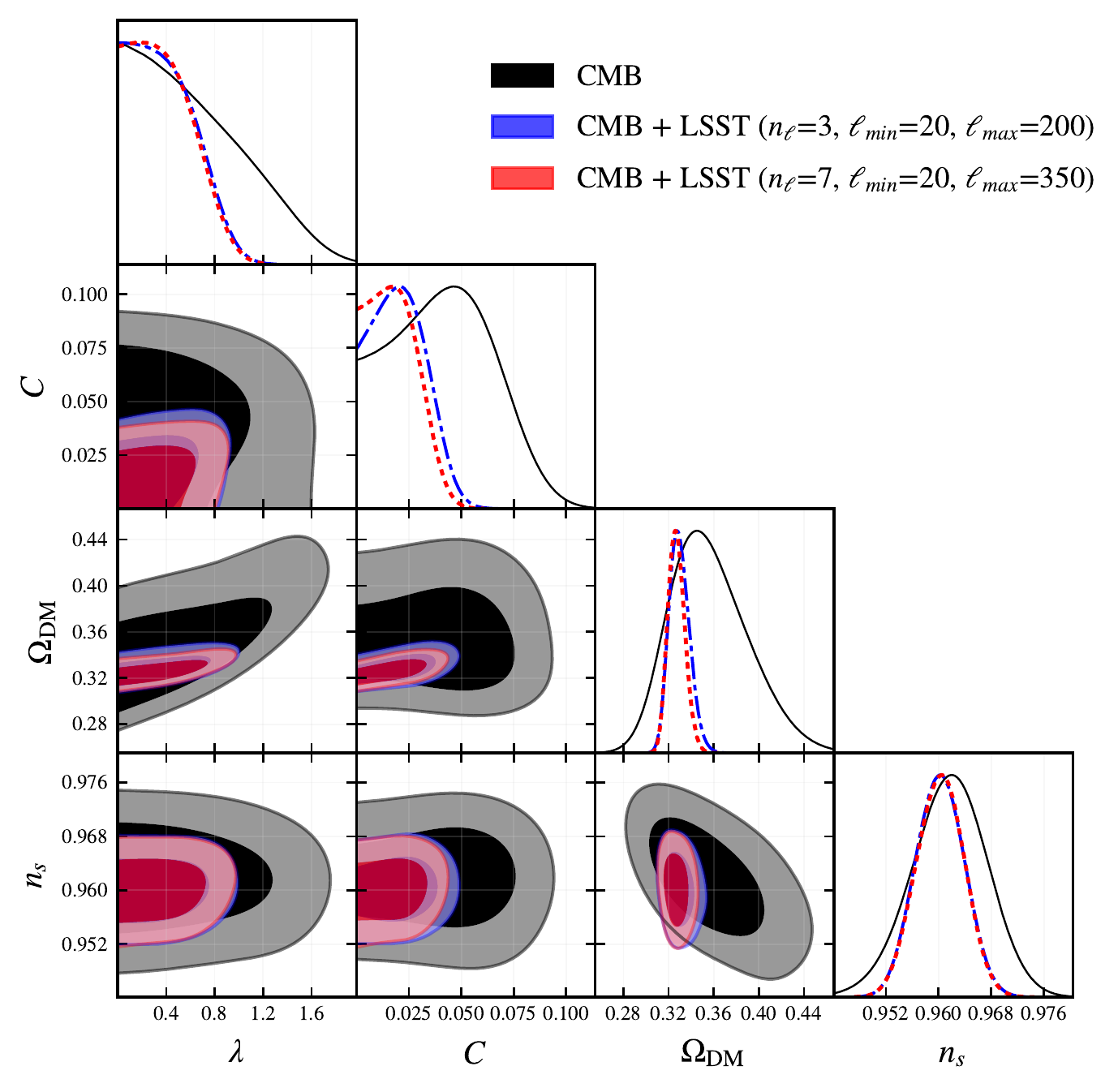}
	\caption{The posterior distribution for cosmological parameters in the context of coupled dark energy models. The CMB data consists of both temperature and polarization data, but excludes lensing reconstruction. This figure shows the final result of combining the CMB and LSST lensing, which can improve upper limits by a factor of two.}
	\label{fig:Plot8} 
\end{figure*}

For qualitative statements at quasi-linear scales and for couplings $|C| \lesssim 0.1$, it is not unreasonable to assume $\Lambda$CDM Halofit with $\Omega_m = \Omega_{\text{DM}}$. Indeed, in~\cite{2016JCAP...01..045C} it was shown that for a limited set of cosmological parameters, couplings in this range introduce only a few percent change in the nonlinear spectrum in comparison to $\Lambda$CDM Halofit at scales $k \lesssim  0.6 \, \text{h/Mpc}$. This approximation is, however, a source of systematic uncertainties in our results that prevents us from stating precisely what would be the ultimate upper limit on the coupling achievable by DES and LSST. We intend to mitigate this systematic in a future work, by generalizing higher order perturbation theory to address coupled dark energy models.

We next turn to the LSST forecast shown in figures~\ref{fig:Plot7} and~\ref{fig:Plot8}. Our LSST chains are not as realistic as the DES forecast, since they do not include photometric redshift biases, but they do better illustrate the correlations explored in this work due to the increased constraining power of LSST compared to DES. They also show that increasing the precision at larger scales, where systematic uncertainties are less severe, can in principle improve the DES constraints by a factor of two. To reach the stringent upper limit of $C < 0.01$ at the two sigma level, knowledge of the power spectrum in the deep nonlinear regime seems to be required. Nevertheless, to provide a definite answer for the achievable LSST upper limit, systematics must be addressed. We postpone this investigation to future work. 


\section{Discussion} \label{discussion}
The possibility of an interacting dark sector, perhaps even mimicking the complexity of the visible sector, has been considered in many studies and is well-motivated both through candidate models of high energy physics and by the considerations of effective field theory. Modern cosmological data allows for constraints on such proposals through the combination of multiple datasets relevant to physics at many different scales. 

In this paper we have revisited a relatively simple realization of this idea, in which a single component of dark matter interacts with a single dark energy field through a coupling that is described by a single dimensionless parameter $C$. Previous work using CMB data has shown that energy transfer from dark matter to dark energy ($C>0$ with our conventions) is preferred at small statistical significance by current observations, mainly because of the lower power in the temperature power spectrum at the large scales observed in the Planck data. Interestingly, the preference for a smaller value of $\Omega_{\text{DM}}$ when CMB data is combined with BAO data slightly increases the posterior for $C \gtrsim 0.05$. However, at the same time, Planck data rules out a coupling $C\gtrsim 0.1$, and we have seen that the addition of low redshift information from both BAO and type IA Supernova does not change this upper limit. Finally, we have observed an emergent correlation between the coupling and the sum of the neutrino masses when CMB and BAO data are combined; finding that higher coupling values relax the upper limit on the neutrino masses.

We have used weak lensing and galaxy clustering in the data forecasts for both the DES and LSST surveys to demonstrate a correlation between the inflationary spectral index and the dark sector coupling, so that, at redshifts probed by large-scale structure the effect of a positive $C$ in the matter power spectrum is similar to that of changing the tilt. When including the CMB data, which fixes $\theta_A$ and probes $n_s$ quite well, the size of the effect of the coupling on the linear power spectrum is exacerbated (see figure~\ref{fig:Plot4}). Therefore, the combination of lensing and clustering of galaxies and CMB data has allowed us to demonstrate an improvement on the constraints on the coupling strength without entering the deeply nonlinear regime.

The tightest constraint on the coupling strength ($C \lesssim 0.03$) arises when combining CMB and LSST data. Further improvement on this constraint could be achieved by better modeling the matter power spectrum deep into the nonlinear regime, but this option requires expensive N-body simulations. Another possibility to push DES and LSST limits even further is to include CMB convergence maps and their correlations with lensing and clustering of galaxies \cite{2016arXiv160701761S}. Nevertheless, we have seen that $C \sim 0.1$ provides almost no improvement in $\chi^2$ with current data. Finally, we have observed that these models are not able to address the $H_0$ and $\sigma_8$ tensions between CMB and low redshift data at the same time. Therefore, we believe that constraints at the level $C \lesssim 0.03$, already diminish significantly the appeal of such models.  
\section{Acknowledgments}
We thank Daniel Grin, Rog\'erio Rosenfeld and Bhuvnesh Jain for helpful discussions. The work of M.C. and M.T was supported in part by NASA ATP grant NNX11AI95G. M.T. was also supported in part by US Department of Energy (HEP) Award DE-SC0013528. Computing resources were provided by the University of Chicago Research Computing Center. VM was supported in part by the Charles E.~Kaufman Foundation, a supporting organization of the Pittsburgh Foundation, and in part by the Penn Center for Particle Cosmology. EK was supported by a Kavli Fellowship at Stanford University.

\appendix
\onecolumngrid
\section{Perturbations in synchronous gauge} \label{ps}
In this Appendix, we compute the linear perturbations for the coupled model. We will follow~\cite{Ma:1995ey} and work in conformal time, in the synchronous gauge, iwith metric is given by
\begin{equation}
\ud s^2=a(\tau)^2\left(-\ud\tau^2+(\delta_{ij}+h_{ij})\ud x^i \ud x^j\right) \ ,
\end{equation}
where the scalar mode of the metric perturbations in Fourier space $k$ is written as 
\begin{equation}
h_{ij}(\mathbf{x},\tau)=\int\ud^3ke^{i\mathbf{k}\cdot\mathbf{x}}\left(\hat{k}_i\hat{k}_jh(\mathbf{k},\tau)+(\hat{k}_i\hat{k}_j-\frac{1}{3}\delta_{ij})\right)6\eta(\mathbf{k},\tau), \quad \mathbf{k}=k\hat{k} \ ,
\end{equation}
and we have introduced the fields $h(\mathbf{k},\tau)$ and $\eta(\mathbf{k},\tau)$. The components of the perturbed energy-momentum  tensor read
\begin{equation}
\begin{split}
T^0_0&=-\rho-\delta\rho \ , \\
T^0_i&=(\rho+P)v_i \ ,\\
T_0^i&=-(\rho+P)v^i \ ,\\
T^i_j&=(\rho+P)\delta^i_j+p\Pi^i_j \ .
\end{split}
\end{equation}
The perturbed Einstein field equations in this gauge are then
\begin{equation}
\begin{split}
	\dot{\eta}_T \Big(1 - \frac{3K}{k^2}\Big) - \frac{K}{2k ^2}\dot{h}_L  &=\frac{\kappa^2 a^2}{2} \sum_J (\rho_J+ p_J) \frac{v_J}{k} \ , \\
	\ddot{h}_L + \mathcal{H} \dot{h}_L &= -\kappa^2 a^2 \sum_J  (\delta \rho_J + 3 \delta p_J) \ , \\
	k^2\Big(1 - \frac{3K}{k^2}\Big)\eta_T - \frac{1}{2}\mathcal{H}\dot{h}_L &= -\kappa^2a^2 \sum_J  \delta \rho_J \ ,
\end{split}
\end{equation}
where $K$ is the spatial curvature and $\kappa\equiv\sqrt{8\pi G}$. When working in synchronous gauge, there is still a residual gauge freedom given by the coordinate transformations
\begin{equation}
\begin{split}
	&\tau\rightarrow\tau+\frac{c_{0}}{a}\mathcal{R}(e^{i\mathrm{k}\cdot \mathrm{x}}) \ ,\\
	&x^{j}\rightarrow x^{j}+kc_{0}\mathcal{R}(i\hat{k}_{j}e^{i\mathrm{k}\cdot \mathrm{x}})\int\frac{d\tau}{a}+c_1 \ ,
\end{split}
\end{equation}
where $c_{0}$ and $c_1$ are constants. These constants will be fixed once we pick the initial conditions, which are computed in detail in Appendix \ref{shp}. 

For the interacting dark energy model studied in this paper, the perturbed Einstein equations become 
\begin{equation}
\begin{split}
	\eta_T -\frac{1}{2}\tilde{\mathcal{H}} \dot{h}_L =&-\frac{ 3 \tilde{\mathcal{H}}^2 }{2}  \big(R_b\delta_b+ R_c \big(\delta_c  +  \alpha' \varphi\big) +R_{\phi} \delta_\phi + R_\gamma \delta_\gamma  + R_\nu\delta_\nu \big) \ , \\
	\dot{\eta}_T  =& \frac{3}{2} \tilde{\mathcal{H}}^2  \Big(R_bV_b+ R_c V_c  + R_{\phi}(1+w_{\phi})  V_\phi + \frac{4}{3}R_\nu V_\nu + \frac{4}{3}R_\gamma V_\gamma  \Big) \ , \\
	\ddot{h}_L + \tilde{\mathcal{H}} \dot{h}_L =&  -3 \tilde{\mathcal{H}}^2  \Big( R_{\phi} \delta_\phi   \Big(1 + 3 \delta p_\phi/\delta\rho_\phi\Big) +R_b\delta_b+ R_c \left(\delta_c  +  \dot{\alpha} k \varphi\Big/\Big(\frac{d \phi}{d\tau}\Big)\right) \\
	&+ 2 R_\nu \delta_\nu + 2 R_\gamma \delta_\gamma \Big) \ ,
\end{split}
\end{equation}
where we have written the perturbed densities as
$\rho_{c}(x,\ \tau)=\rho_{i}(\tau)(1+\delta_{i}(x,\ \tau))$, and defined  $R_i=\rho_i\big/\rho_\text{total}$, $\tilde{H} = H/k$, $'\equiv\tfrac{\ud}{\ud \phi}$, and $\dot{}\equiv d\big/d x$, with $x\equiv k\tau$. Meanwhile, the equations of motion for the dark sector fields are given by
\begin{equation}
\begin{split}
	\ddot{\varphi} +\frac{1}{2}\dot{h}_L \dot{\phi}+ 2 \mathcal{H}\dot{\varphi} +  \varphi \Big(k^2 + a^2 V''+ a^2 e^{\alpha} \rho_c \big(\alpha'^2  + \alpha''\big) \Big)  &=  - a^2 e^{\alpha} \alpha' \rho_c \delta_c \ , \\
	\dot{\delta}_c + V_c + \frac{1}{2} \dot{h}_L &= 0 \ , \\
	k \dot{V}_c + \big( \mathcal{H} + \dot{\phi} \alpha' \big) k V_c &= k^2 \alpha' \varphi \ .
\end{split}
\end{equation}

We next focus on the perturbation equations for baryons, photons and neutrinos. When computing the initial conditions, we ignore higher order moments in the Boltzmann hierarchy of the neutrinos, since they would be suppressed by additional powers of $k\tau$, and will not be relevant for our discussion~\cite{Ma:1995ey}. Given this, the perturbed equations of motion are given by
\begin{equation}
\begin{split}
\dot{\delta}_\nu  +\frac{4}{3} V_\nu + \frac{2}{3}\dot{h}_L&=0 \ ,\\
\dot{V}_\nu -  \frac{\delta_\nu}{4} + \sigma_\nu&=0 \ ,\\
\dot{\sigma}_\nu - \frac{4}{15}  V_\nu + \frac{3 F_\nu^{(3)}}{10}- \frac{2}{15} \dot{h}_L - \frac{4}{5}\dot{\eta}_T &=0 \ ,\\
\dot{F}_\nu^{(3)} - \frac{6 \sigma_\nu}{7}&=0 \ ,
\end{split}
\end{equation}
and
\begin{equation}
\begin{split}
\dot{\delta}_b + V_b + \frac{1}{2} \dot{h}_L &= 0 \ , \\
k^2 \dot{V}_b +\mathcal{H} k V_b &=-\frac{4 \rho_\gamma}{3  \rho_b}a n_e \sigma_T k (V_b-V_\gamma) \ ,  \\
\dot{\delta}_\gamma+ \frac{4}{3}V_\gamma + \frac{2}{3}\dot{h}_L &= 0 \ , \\
\dot{V}_\gamma - \frac{1}{4}\delta_\gamma &=\frac{1}{k}a\, n_e \sigma_T  (V_b-V_\gamma) \ .  \\
\end{split}
\end{equation}
We take the tight-coupling approximation, since the baryons and photons are strongly coupled due to Thomson scattering. In this approximation the velocity perturbations are $V_\gamma = V_b \equiv V_{\gamma b}$. This implies that the equation for the baryon-photon fluid velocity perturbation reads
\begin{equation}
\left(1+\frac{4 \rho_\gamma}{3  \rho_b} \right) \dot{V}_{\gamma b} =-\frac{\mathcal{H}}{k} V_{\gamma b} + \frac{4 \rho_\gamma}{3  \rho_b}\frac{\delta_\gamma}{4} \ .
\end{equation}

\section{Super-horizon perturbations} \label{shp}
The CAMB code begins mode integration well outside the horizon, and we therefore seek super-horizon initial conditions. We use a series solution method, redefine the perturbation variables, and expand the background functions in powers of $x= k\tau$ \cite{Ma:1995ey,Bucher:1999re}. This amounts to an early-time and super-horizon expansion. The new variables are
\begin{eqnarray}
\begin{split}
\Theta_h = \dot{h}_L,  \quad
	\tilde{\delta}_i = \frac{\delta_i}{x},  \quad
	\tilde{\sigma}_\nu = \frac{\sigma_\nu}{x},  \quad
	\tilde{V}_i = \frac{V_i}{x^2},  \quad
	\tilde{F}_\nu =\frac{{F}_\nu}{x^2},  \quad
	\tilde{\varphi}=\frac{\varphi}{x},   \\
	\tilde{u}_\phi=\frac{\ud \tilde{\varphi}}{ \ud\ln x},\quad
	\tilde{R}_c\frac{\rho_c}{\rho_m},\quad 
	\tilde{R}_b=\frac{\rho_b}{\rho_m},\quad  
	\tilde{R}_\nu=\frac{\rho_\nu}{\rho_r},\quad 
	\tilde{R}_\gamma=\frac{\rho_\gamma}{\rho_r}.
	\end{split}
\end{eqnarray}
With these definitions, the perturbed equations read
\begin{align*}
	\frac{d\tilde{\delta}_\gamma}{d\ln x} &= - \tilde{\delta}_\gamma -\frac{4}{3} x^2 \tilde{V}_{\gamma b} - \frac{2}{3} \Theta_h \ ,\\
	\frac{d\tilde{\delta}_\nu}{d\ln x} &= - \tilde{\delta}_\nu -\frac{4}{3} x^2 \tilde{V}_\nu - \frac{2 }{3}\Theta_h \ , \\
	\frac{d\tilde{\delta}_c}{d\ln x} &= - \tilde{\delta}_c - x^2 \tilde{V}_c - \frac{1}{2}\Theta_h \ , \\
	\frac{d\delta_b}{d\ln x} &= - \tilde{\delta}_b - x^2 \tilde{V}_{\gamma b} - \frac{1}{2}\Theta_h \ , \\
	\frac{d \tilde{\varphi}}{d\ln x }  &=\tilde{u}_\phi \ ,
	\end{align*}
	\begin{align*}
	\frac{d\tilde{V}_{\gamma b}}{d\ln x} &\equiv  -2\tilde{V}_{\gamma b}+\left(1-\frac{3}{4}\frac{\tilde{R}_b}{\tilde{R}_\gamma}\left(m_1 \frac{x}{k} +m_2 \left(\frac{x}{k}\right)^2\right)\left(1-r_1 \frac{x}{k} -r_2 \left(\frac{x}{k}\right)^2\right)\right)\times\\
	&\times\left[ \frac{3}{4}\frac{\tilde{R}_b}{\tilde{R}_\gamma}\left(m_1 \frac{x}{k} +m_2 \left(\frac{x}{k}\right)^2\right)\left(1-r_1 \frac{x}{k} -r_2 \left(\frac{x}{k}\right)^2\right)\big(1 + h_0 \frac{x}{k} +h_1 \left(\frac{x}{k}\right)^2  \big) \tilde{V}_{\gamma b} + \frac{\tilde{\delta}_\gamma}{4}\right] \ ,\\
	\frac{d \tilde{V}_\nu}{d\ln x }  &=  -2 \tilde{V}_\nu + \frac{1}{4}\tilde{\delta}_\nu - \tilde{\sigma}_\nu \ , \\
	\frac{\ud \tilde{V}_c}{\ud \ln x }  &=  - \Big(3 + h_0 \frac{x}{k} +h_1 \left(\frac{x}{k}\right)^2\Big)  \tilde{V}_c + \alpha' (\tilde{\varphi} -\frac{\ud \phi}{\ud \ln x}\tilde{V}_c) \ , \\
	\frac{\ud \tilde{u}_\phi}{\ud \ln x }&=-2(\frac{3}{2}+h_0 \frac{x}{k} +h_1 \left(\frac{x}{k}\right)^2)\tilde{u}_\phi-x^2\left[1+a_1^2\left(\frac{x}{k}\right)^2\left(1+a_2\frac{x}{k}\right)^2\frac{V''}{k^2}\right]\tilde{\varphi}-\frac{1}{2}\Theta_h \frac{\ud \phi}{\ud \ln x}\nonumber\\
	&+\left[3\left(1+h_0 \frac{x}{k} +h_1 \left(\frac{x}{k}\right)^2\right)\left(m_1 \frac{x}{k} +m_2 \left(\frac{x}{k}\right)^2\right)\tilde{R}_c\left(\alpha'^2+\alpha''\right)-2(1+h_0 \frac{x}{k} +h_1 \left(\frac{x}{k}\right)^2)\right]\tilde{\varphi} \nonumber \nonumber\\
	&-3\alpha'\tilde{R}_c\tilde{\delta}_c\left(1+h_0 \frac{x}{k} +h_1 \left(\frac{x}{k}\right)^2\right)^2\left(m_1 \frac{x}{k} +m_2 \left(\frac{x}{k}\right)^2\right) \ , \nonumber \nonumber
		\end{align*}
	\begin{align}
	\frac{d\tilde{\sigma}_\nu}{d\ln x} &= -\tilde{\sigma}_\nu - \frac{3 x^2 \tilde{F}_\nu^{(3)}}{10} +  \frac{4}{15} x^2 \tilde{V}_\nu + \frac{2}{15} \Theta_L
	+\frac{2}{5}\frac{x}{k}\frac{\ud \phi}{\ud \ln x}\tilde{\varphi}+\left(1+h_0 \frac{x}{k} +h_1 \left(\frac{x}{k}\right)^2\right)^2\times \nonumber\\
	&\times\left[ \frac{8}{5}\big(  \tilde{R}_\nu V_\nu + \tilde{R}_\gamma V_{\gamma b} \big) \left(1+r_1 \frac{x}{k} +r_2 \left(\frac{x}{k}\right)^2\right)+ \frac{6}{5}\big(  \tilde{R}_c V_c + \tilde{R}_b V_{\gamma b} \big) \left(m_1 \frac{x}{k} +m_2 \left(\frac{x}{k}\right)^2\right)\right] \ ,\nonumber  \\
	\frac{d \tilde{F}_\nu^{(3)}}{d\ln x} &= -2 \tilde{F}_\nu^{(3)} + \frac{6}{7} \tilde{\sigma}_\nu \ ,\nonumber \\
	\frac{d\Theta_h}{d\ln x} &= -\left(1+h_0 \frac{x}{k} +h_1 \left(\frac{x}{k}\right)^2\right)\Theta_h-4 \frac{\ud \phi}{\ud \ln x} \tilde{u}_\phi+2\,a_1\left(\frac{x}{k}\right)^3\left(1+a_2\frac{x}{k}\right)V'\tilde{\varphi}\nonumber \\
	& -\left(1+h_0 \frac{x}{k} +h_1 \left(\frac{x}{k}\right)^2\right)^2\Biggl[6  \left( \tilde{R}_\nu \tilde{\delta}_\nu + \tilde{R}_\gamma \tilde{\delta}_\gamma \right) \left(1+r_1 \frac{x}{k} +r_2 \left(\frac{x}{k}\right)^2\right)\nonumber \\
	&-3  \left( \tilde{R}_c\left(\tilde{\delta}_c+\alpha' \tilde{\varphi}\right)  + \tilde{R}_b \tilde{\delta}_b \right) \left(m_1 \frac{x}{k} +m_2 \left(\frac{x}{k}\right)^2\right) \Biggl] \ ,\nonumber\\
	\frac{d \eta_T}{d\ln x}  &= \left(1+h_0 \frac{x}{k} +h_1 \left(\frac{x}{k}\right)^2\right)^2\Biggl[2 x  \left(  \tilde{R}_\nu \tilde{V}_\nu + \tilde{R}_\gamma \tilde{V}_{\gamma b}  \right) \left(1+r_1 \frac{x}{k} +r_2 \left(\frac{x}{k}\right)^2\right) \nonumber\\
	&+ \frac{3x}{2} \left(\tilde{R}_c \tilde{V}_c + \tilde{R}_b \tilde{V}_{\gamma b} \right) \left(m_1 \frac{x}{k} +m_2 \left(\frac{x}{k}\right)^2\right)\Biggl]+\frac{1}{2}\frac{x}{k}\frac{\ud \phi}{\ud \ln x}\tilde{\varphi} \ ,
\end{align}
where
\begin{eqnarray}
\begin{split}
	a_1=\sqrt{\frac{\rho_r^0}{3}} \ ,\quad a_2=\sqrt{\frac{3}{\rho_r^0}}  \frac{\rho_m^o}{12} \ ,\quad h_0=\frac{\rho_m^0}{4 \sqrt{3\rho_r^0}} \ , \quad h_1=-\frac{(\rho _m^0)^2}{48 \rho_r^0} \ ,\\
	m_1=-r_1=\frac{\rho _m^0}{\sqrt{3\rho_r^0}} \ ,\quad m_2=-r_2=-\frac{(\rho _m^0)^2}{4 \rho_r^0} \ .
	\end{split}
\end{eqnarray}
To obtain these expansions, we have used the Friedmann equation 
\begin{equation}
	3M_{\text{pl}}^2\left(\frac{\ud a}{\ud \tau}\right)^{2}=\rho_{r}^0+a\rho_{b}^0+ae^{\alpha(\phi)}\rho_{c}^0+a^4\rho_\phi \ ,
\end{equation}
where we have fixed the scale factor today to be $a(\tau_0)=1$. At early times we have $a\sim x/k$, and by writing $
\rho_\phi\propto x^n$, we see that the dark energy density term can be neglected at order $x^2$ if $n\geq-1$. Furthermore, we expand the conformal coupling in a Taylor series around $\tau=0$
\begin{equation}
	e^{\alpha(\phi)}=e^{\alpha(\phi(\tau_0))}+\alpha'\frac{\ud \phi}{\ud \tau}e^{\alpha(\phi(\tau_0))}\tau \ .
\end{equation}
Given that at early times $\rho_\phi\simeq (d \phi \big/d \tau)^2\big/a^2$, the second term is of order $\tau^{n/2+2}$ and can be neglected if $n\gtrsim0$ (as long as $\alpha'e^{\alpha(\phi(\tau_0))}\lesssim\mathcal{O}(\tau^0)$). In addition, since 
\begin{equation}
	\frac{\ud \phi}{\ud \ln x}=\frac{x}{k}\frac{\ud \phi}{\ud \tau}\simeq\left(\frac{x}{k}\right)^{n/2+2} \ ,
\end{equation}
we may neglect any term that contains $d \phi \big/ d \ln x$.

The term containing the scalar field potential can also be neglected as long as $V''(\tau_0)<1/x^4$, which is easily satisfied by the exponential potential. The terms $(x \big/ k)\tilde{R}_c f(\alpha)$, where $f(\alpha)$ is either $\alpha'$, $(\alpha')^2$, or $\alpha''$, should be included as long as $\tilde{R}_c f(\alpha)\gtrsim 1$. The inclusion of these terms, even if they are smaller than unity, would not change the results at next to leading order. Given this, we proceed to include them in order to explore the whole parameter space consistently. To sum up, the assumptions we make are 
\begin{equation}
	\rho_\phi\lesssim x^0,\quad V''(\tau_0)<1/x^4 \ .
\end{equation}
We have checked and confirmed the validity of these assumptions against the numerical solution for the background equations in the parameter space defined by $0<\lambda<\sqrt{3}$ and $|C|<\sqrt{3}\big/2$. 

In order to solve this system of equations, we write the equations of motion in the form
\begin{align}
	\frac{d \vec{u}}{d\ln x} = (A_0 + A_1 x + A_2 x^2 + A_3 x^3....) \vec{u} \ ,
\end{align}
with the vector $\vec{u}$ given by
\begin{align}
	\vec{u}^T = \big\{\tilde{\delta}_\gamma, \tilde{\delta}_\nu, \tilde{\delta}_c, \tilde{\delta}_b,  \tilde{\varphi}, \tilde{V}_{\gamma b},  \tilde{V}_\nu,  \tilde{V}_c, \tilde{u}_\phi, \tilde{\sigma}_\nu, \tilde{F}_\nu^{(3)},    \Theta_h, \eta_T \big\} \ .
\end{align}
The lowest order corrections are then found by solving the system
\begin{equation}
\begin{split}
	(A_0 - \alpha \mathcal{I} ) \vec{u}_0 &= 0 \ , \\
	( (\alpha+1) \mathcal{I} - A_0) \vec{u}_1  &= A_1 \vec{u}_0 \ ,
\end{split}
\end{equation}
where $\alpha$ is the eigenvalue corresponding to the mode $\vec{u}_0$. Using this procedure, we find that the expansion for the adiabatic mode at next to leading order is
\begin{equation}
\begin{split}
	&\text{ADIABATIC MODE} \\
\delta_\gamma&=-\frac{x^2}{3}+\frac{4 R_{\text{m-r}}}{15 k}x^3+\mathcal{O}(x^4) \ ,\\
\delta_\nu&=-\frac{x^2}{3}+\frac{4 R_{\text{m-r}}}{15 k}x^3+\mathcal{O}(x^4) \ ,\\
\delta_c&=-\frac{x^2}{4}+\frac{R_{\text{m-r}}}{5 k}x^3+\mathcal{O}(x^4) \ ,\\
\delta_b&=-\frac{x^2}{4}+\frac{R_{\text{m-r}}}{5 k}x^3+\mathcal{O}(x^4) \ ,\\
\varphi&=\frac{R_{\text{m-r}}\alpha'\tilde{R}_c}{16 k}x^3+\mathcal{O}(x^4) \ , \\
V_{\gamma b}&=-\frac{1}{36}x^3+\frac{(5
	\tilde{R}_c+8\tilde{R}_\nu-13) R_{\text{m-r}}}{480 k (\tilde{R}_\nu-1)}x^4+\mathcal{O}(x^5) \ ,\\
V_{\nu}&=-\frac{4 \tilde{R}_\nu+23}{144 \tilde{R}_\nu+540}x^3+\frac{\left(1475-5 \tilde{R}_c (4 \tilde{R}_\nu+15)+4 \tilde{R}_\nu (8 \tilde{R}_\nu+115)\right) R_{\text{m-r}}}{240 k (2
	\tilde{R}_\nu+15) (4 \tilde{R}_\nu+15)}x^4+\mathcal{O}(x^5) \ ,\\
V_{c}&=\frac{R_{\text{m-r}}\alpha'^2\tilde{R}_c}{8 k}x^4+\mathcal{O}(x^5) \ , \\
\dot{\varphi}&=\frac{3 R_{\text{m-r}} \alpha' \tilde{R}_c}{16 k}x^2+\mathcal{O}(x^3) \ ,\\
\sigma_\nu&=\frac{2}{45+12\tilde{R}_\nu}x^2+\frac{(4 \tilde{R}_\nu(\tilde{R}_c-5)+5(3\tilde{R}_c-23)) R_{\text{m-r}}}{12 k (2 \tilde{R}_\nu+15) (4 \tilde{R}_\nu+15)}x^3+\mathcal{O}(x^4) \ ,\\
F_\nu&=\frac{4}{84 \tilde{R}_\nu+315}x^3++\frac{(4 \tilde{R}_\nu(\tilde{R}_c-5)+5(3\tilde{R}_c-23))R_{\text{m-r}}}{56 k (2 \tilde{R}_\nu+15) (4 \tilde{R}_\nu+15)}x^4+\mathcal{O}(x^5) \ ,\\
h&=\frac{k}{2}\tau^2-\frac{2 R_{\text{m-r}}}{5 k}x^3+\mathcal{O}(x^4) \ ,\\
\eta&=1-\frac{5+4\tilde{R}_\nu}{180+48\tilde{R}_\nu}x^2+\mathcal{O}(x^3) \ ,
\end{split}
\end{equation}
where we have defined $R_{\text{m-r}}\equiv\rho _m^0\big/\sqrt{3\rho_r^0}$. From this, we see that at leading order there are no contributions to the dark matter velocity, to $\eta$, or to the field variables. In order to compare our results with~\cite{Amendola:1999dr}, where initial conditions for this kind of model were obtained, it is important to note that  this paper uses a different definition of the dark matter density. In~\cite{Amendola:1999dr} $\rho_c$ and $\delta_c$ are defined as the coupled quantities, that is, $\rho_c^\text{that paper}=e^{\alpha(\phi)}\rho_c^\text{this paper}=\bar\rho_c^\text{ this paper}$.

We also obtain the isocurvature modes corresponding to the interacting dark sector. These modes read
\begin{align}
	&\text{DARK ENERGY ISOCURVATURE} \nonumber\\
	\delta_\gamma&=-\frac{2 R_{\text{m-r}}\tilde{R}_c\alpha'}{3 k}x+\mathcal{O}(x^2) \ ,\nonumber\\
	\delta_\nu&=-\frac{2 R_{\text{m-r}}\tilde{R}_c\alpha'}{3 k}x+\mathcal{O}(x^2)\ ,\nonumber\\
	\delta_c&=-\frac{R_{\text{m-r}}\tilde{R}_c\alpha'}{2 k}x+\mathcal{O}(x^2) \ ,\nonumber\\
	\delta_b&=-\frac{R_{\text{m-r}}\tilde{R}_c\alpha'}{2 k}x+\mathcal{O}(x^2) \ ,\nonumber\\
	\varphi&=-1-\frac{R_{\text{m-r}}(1+3\tilde{R}_c(\alpha'^2+\alpha''))}{4 k}x+\mathcal{O}(x^2) \ , \nonumber\\
	V_{\gamma b}&=-\frac{R_{\text{m-r}}\tilde{R}_c\alpha '}{12 k}x^2+\mathcal{O}(x^3) \ ,\nonumber\\
	V_{\nu}&=-\frac{4 \tilde{R}_\nu+13}{12( 4\tilde{R}_\nu+5)}\frac{R_{\text{m-r}}\tilde{R}_c\alpha'}{k}x^2+\mathcal{O}(x^3) \ ,\nonumber\\
	V_{c}&=-\frac{\alpha'}{2}x-\frac{R_{\text{m-r}}\alpha'}{24k}\left(1+3\tilde{R}_c(\alpha'^2+\alpha'')\right)x^2+\mathcal{O}(x^3) \ , \nonumber\\
	\dot{\varphi}&=-\frac{R_{\text{m-r}}(1+3\tilde{R}_c(\alpha'^2+\alpha''))}{4 k}+\mathcal{O}(x) \ ,\nonumber\\
	\sigma_\nu&=-\frac{3}{5+4\tilde{R}_\nu}\frac{R_{\text{m-r}}\tilde{R}_c\alpha '}{ k}x+\mathcal{O}(x^2) \ ,\nonumber\\
	F_\nu&=-\frac{9}{35+28\tilde{R}_\nu}\frac{R_{\text{m-r}}\tilde{R}_c\alpha '}{ k}x^2+\mathcal{O}(x^3) \ ,\nonumber\\
	h&=-\frac{R_{\text{m-r}}\tilde{R}_c\alpha '}{ k}x+\mathcal{O}(x^2) \ ,\nonumber\\
	\eta&=\mathcal{O}(x).
\end{align}

\begin{align}
	&\text{CDM ISOCURVATURE} \nonumber\\
	\delta_\gamma&=\frac{2 R_{\text{m-r}}\tilde{R}_c}{3 k}x+\mathcal{O}(x^2) \ ,\nonumber\\
	\delta_\nu&=\frac{2 R_{\text{m-r}}\tilde{R}_c}{3 k}x+\mathcal{O}(x^2) \ ,\nonumber\\
	\delta_c&=1+\frac{R_{\text{m-r}}\tilde{R}_c}{2 k}x+\mathcal{O}(x^2) \ ,\nonumber\\
	\delta_b&=\frac{R_{\text{m-r}}\tilde{R}_c}{2 k}x+\mathcal{O}(x^2) \ ,\nonumber\\
	\varphi&=-\frac{3 R_{\text{m-r}}\alpha'\tilde{R}_c}{2 k}x+\mathcal{O}(x^2) \ , \nonumber
		\end{align}
	\begin{align}
	V_{\gamma b}&=\frac{R_{\text{m-r}}\tilde{R}_c}{12 k}x^2+\mathcal{O}(x^3) \ ,\nonumber\\
	V_{\nu}&=\frac{R_{\text{m-r}}\tilde{R}_c}{12 k}x^2+\mathcal{O}(x^3) \ ,\nonumber\\
	V_{c}&=-\frac{R_{\text{m-r}}\alpha'^2\tilde{R}_c}{2 k}x^2+\mathcal{O}(x^3) \ , \nonumber\\
	\dot{\varphi}&=-\frac{3R_{\text{m-r}}\alpha'\tilde{R}_c}{2k}+\mathcal{O}(x) \ ,\nonumber\\
	\sigma_\nu&=\mathcal{O}(x^2) \ ,\nonumber\\
	F_\nu&=\mathcal{O}(x^3) \ ,\nonumber\\
	h&=\frac{R_{\text{m-r}}\tilde{R}_c}{ k}x+\mathcal{O}(x^2) \ ,\nonumber\\
	\eta&=\mathcal{O}(x) \ .
\end{align}
\bibliography{DEDM}
\end{document}